\documentclass[aps,prd,nofootinbib,twocolumn,floatfix,letterpaper,superscriptaddress,showpacs]{revtex4}
\usepackage{graphicx}\usepackage{natbib}\usepackage{amsmath}
\pdfoutput=1
\usepackage{color}
\definecolor{c1}{rgb}{1.0,0.188,0.188}
\definecolor{c3}{rgb}{0.222, 0.667, 0.798}
\definecolor{c4}{rgb}{0.95, 0.361, 0.011}
\definecolor{c2}{rgb}{0.568, 0.603, 0.568}

\begin{document}
 
\title{Cosmic PeV Neutrinos and the Sources of Ultrahigh Energy Protons}

\author{Matthew D. Kistler}
%\email{mdkistler@lbl.gov}
\affiliation{Lawrence Berkeley National Laboratory and Department of Physics, University of California, Berkeley, CA 94720}
\affiliation{Kavli Institute for Particle Astrophysics and Cosmology, Stanford University, SLAC National Accelerator Laboratory, Menlo Park, CA 94025}
\affiliation{Einstein Fellow}

\author{Todor Stanev}
\affiliation{Bartol Research Institute, Department of Physics and Astronomy, University of Delaware, Newark, DE 19716}

\author{Hasan Y{\"u}ksel}
\affiliation{Theoretical Division, Los Alamos National Laboratory, Los Alamos, NM 87544}
\affiliation{Department of Physics, Mimar Sinan Fine Arts University, Bomonti 34380, \.{I}stanbul, Turkey}

\date{October 31, 2014}

\begin{abstract}
The IceCube experiment recently detected the first flux of high-energy neutrinos in excess of atmospheric backgrounds.  We examine whether these neutrinos originate from within the same extragalactic sources as ultrahigh-energy cosmic rays.  Starting from rather general assumptions about spectra and flavors, we find that producing a neutrino flux at the requisite level through pion photoproduction leads to a flux of protons well below the cosmic-ray data at $\sim\!10^{18}\,$eV, where the composition is light, unless pions/muons cool before decaying.  This suggests a dominant class of accelerator that allows for cosmic rays to escape without significant neutrino yields.
\end{abstract}
%

%\preprint{UCB-NPAT-12-027, NT-LBNL-12-031,LA-UR-12-26976}

% 95.85.Ry     Neutrino, muon, pion, and other elementary particles; cosmic rays
% 98.70.Rz     gamma-ray sources; gamma-ray bursts
% 98.70.-f	        Unidentified sources of radiation outside the Solar System
% 98.70.Sa     Cosmic rays (including sources, origin, acceleration, and interactions)
% 98.35.Eg     Electric and magnetic fields of the Milky Way galaxy
%,showpacs
\maketitle

%--------------------------------------------------------------------%
\section{Introduction}
High-energy astrophysical neutrinos have much to tell us about the most extreme environments in the Universe; however, finding them is a difficult endeavor \cite{Gaisser:1994yf,Learned:2000sw,Halzen:2002pg,Becker:2007sv}.  Colossal detectors are required \cite{Roberts:1992re,Halzen:1988wr,Barwick:1991ur}, such as IceCube \cite{Ahrens:2002dv}, that can observe the tracks of muons produced in $\nu_\mu$ charged-current scattering or showers (cascades) arising from a variety of channels (as we discuss later).  The first observation of two PeV-energy shower events \cite{Aartsen:2013bka} and numerous $\sim\,$100~TeV events \cite{Whitehorn,Klein:2013} by IceCube from 2010-2013 may represent the discovery of such neutrinos, as atmospheric PeV neutrino fluxes are low \cite{Gaisser:2002jj,Enberg:2008te}.

A likely astrophysical mechanism is pion photoproduction by protons on a photon background, $p\, \gamma\! \rightarrow \! N\,\pi$, leading to neutrinos via the pion decay chain.  An example of this process is the well-known suppression of ultrahigh-energy cosmic-ray (UHECR) proton fluxes at $\gtrsim\,$$10^{19.5}\,$eV due to the cosmic microwave background (CMB), the GZK effect \cite{Greisen:1966jv,Zatsepin:1966jv}.  The measured UHECR spectrum displays a marked downturn near this energy \cite{Abbasi:2007sv,Abraham:2008ru,Abreu:2011pj,AbuZayyad:2012ru}.  However, the resulting $\gtrsim\,$$10^{18}\,$eV neutrinos \cite{Beresinsky:1969qj,Stecker:1978ah,Hill:1983mk,Yoshida:1993pt,Waxman:1998yy,Engel:2001hd} are far too energetic to explain the IceCube events.

The proton energy threshold for pion photoproduction on the cosmic infrared/optical background is lower, leading to lower-energy neutrinos \cite{Stanev:2004kz,Bugaev:2004xt,DeMarco:2005kt}.  However, {\it Fermi} measurements of gamma-ray absorption now indicate a low level of the $\lesssim\,$10~eV diffuse photons \cite{Ackermann:2012} needed to scatter off $\lesssim\,$10$^{17}\,$eV protons to yield $\sim\,$$10^{15}\,$eV neutrinos.  Producing such a cosmogenic neutrino flux at the required level in the 0.1--1~PeV range would generally overproduce the isotropic gamma-ray background~\cite{Roulet}.  This suggests that the PeV neutrinos arose from within some population of sources.  The lack of correlation with the Galactic plane favors an extragalactic origin \cite{Whitehorn,Klein:2013}, with IceCube limits disfavoring GRBs \cite{Abbasi:2012zw,Hummer:2011ms,Cholis:2012kq,Liu:2012pf,Baerwald:2014zga}.

%
%%%%%%%%%%%%%%%%%%%%%%%%%
\begin{figure}[b!]
\vspace{-0.2in}
\includegraphics[width=3.35in,clip=true]{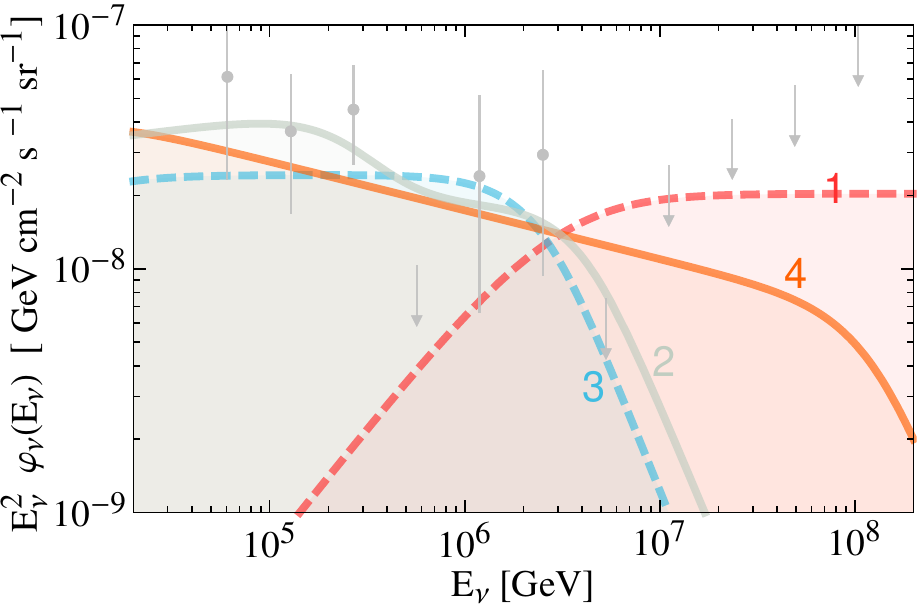}
\vspace{-0.15in}
\caption{Total $\nu \!+\! \bar{\nu}$ fluxes of our four fiducial models.  Normalizations are fixed to yield three $\gtrsim\,$1~PeV events in IceCube for Model {\color{c1} 1}, or ten $\gtrsim\,$100~TeV events for Models {\color{c2} 2}, {\color{c3} 3}, and {\color{c4} 4}, following the methods in the text.  To rescale for the flux of any $\nu$ or $\bar{\nu}$ flavor $i$ multiply by the corresponding post-oscillation $N_i$ divided by $N_{\nu+\bar{\nu}}$ for each model in Table~\ref{table:ratios}.  The IceCube data ({\it circles}) assume a 1:1:1 flavor ratio and $\nu \!=\! \bar{\nu}$ \cite{Klein:2013}.
\label{fluxes}}
\end{figure}
%%%%%%%%%%%%%%%%%%%%%%%%%
%

A long-standing hope is to determine the UHECR sources and ascertain the acceleration mechanism \cite{Hillas:1985is}.  Our goal is to discern what the IceCube neutrinos reveal about the sources of UHECR protons, in particular, whether the neutrinos share a common origin with UHECR in the $\sim\! 10^{18}\,$eV range where the composition is inferred to be light \cite{Abbasi:2004nz,Abbasi:2009nf,Abraham:2010yv,Apel:2013dga,Barcikowski:2013nfa}.  Of special interest are scenarios in which the magnetic fields required to contain protons during their acceleration do not allow for escape prior to energy loss.  Along with pions, photohadronic interactions produce neutrons that may freely leave and later decay into protons.  We use the level of neutrinos implied by IceCube data along with basic suppositions about the means of neutrino production to construct models (see Fig.~\ref{fluxes}) to illustrate necessary properties of these sources, including whether the neutron mechanism is responsible for releasing a sufficient proton flux.

%---------------------------------------------------------------------------%
\begin{table*}[t!]
\caption{Number of neutrinos and antineutrinos in our four models: total $N_{\nu+\bar{\nu}}$ or by flavor $N_i$ at birth (after oscillations).\label{table:ratios}}
\begin{ruledtabular}
\begin{tabular}{lccccccc}
Model		& $N_{\nu+\bar{\nu}}$	&$N_{\nu_e}$	&$N_{\nu_\mu}$	&$N_{\nu_\tau}$	&$N_{\bar{\nu}_e}$	&$N_{\bar{\nu}_\mu}$	&$N_{\bar{\nu}_\tau}$ \\
\hline
{\color{c1} 1}	& 6					& 1 (1.01)		& 2 (1.04)		& 0 (0.95)		& 1 (1.01)		& 2 (1.04)			& 0 (0.95)	\\
{\color{c2} 2} ($\pi$)	& 1				& 0 (0.23)		& 1 (0.40)		& 0 (0.37)		& 0 (0)		& 0 (0)			& 0 (0) \\
{\color{c2} 2} ($\mu$)	& 2			& 1 (0.55)		& 0 (0.23)		& 0 (0.22)		& 0 (0.23)		& 1 (0.40)			& 0 (0.37) \\
{\color{c3} 3}	& 6					& 1 (1.01)		& 2 (1.04)		& 0 (0.95)		& 1 (1.01)		& 2 (1.04)			& 0 (0.95)	\\
{\color{c4} 4}	& 3					& 1 (0.78)		& 1 (0.64)		& 0 (0.58)		& 0 (0.23)		& 1 (0.40)			& 0 (0.37)	
\end{tabular}
\end{ruledtabular}
\end{table*}
%---------------------------------------------------------------------------%

%--------------------------------------------------------------------%
\section{Neutrino Events in IceCube}
The rate of neutrino interactions within a detector directly depends upon the impinging flux along with the odds of a given neutrino interacting, which is based upon the number of target particles and the various scattering cross sections.  In IceCube, at the energies of interest here, the relevant targets are nucleons (protons or neutrons) and electrons.  The wide variety of interactions break down into two broad events classes: those producing a long-ranging muon and those that do not.

When no muon is produced, the energy loss lengths of the products are relatively short, yielding quasi-spherical objects referred to as showers.  While the angular resolution of such events is modest ($\sim\!10^\circ$; \cite{Whitehorn}) the rapid energy deposition can be nearly calorimetric.  On the other hand, muons with multi-TeV energies lose energy over km-scale distances, so that they can even be detected via Cherenkov radiation even when originating well beyond the instrumented detector volume.  Long tracks within the detector allow angular resolutions of $\sim\!1^\circ$, although the lower energy loss rates tend to obscure the initial energy, especially for muons beginning outside the detector.

The great recent advance in this field has been the observation by IceCube first of two $\sim\,$PeV-energy shower events \cite{Aartsen:2013bka}, followed by searches including lower energies yielding 37 total events with deposited energy exceeding 30~TeV, 9 of which display outgoing muon tracks \cite{Whitehorn,Klein:2013}.  The background expectation is $6.6^{+5.9}_{-1.6}$ events from conventional atmospheric neutrinos and $8.4 \pm 4.2$ due to atmospheric muons \cite{Klein:2013}.  The atmospheric neutrino flux from the decays of charmed mesons falls less steeply with energy, but also appears insufficient \cite{Klein:2013}, thus allowing a variety of new astrophysical diagnostics \cite{Pakvasa:2012db,Kalashev:2013vba,Arsene:2013nca,Fox:2013oza,Vissani:2013iga,Laha:2013lka,Murase:2013rfa,Anchordoqui:2013qsi,Winter:2013cla,Ahlers:2013xia,Halzen:2013dva,Anchordoqui:2013dnh,Fang:2014uja,Kachelriess:2014oma,Learned:2014vya,Winter:2014pya}.

We present an analytical method useful for examining the extent to which the neutrinos giving rise to these excess events may have originated from the sources of ultrahigh-energy cosmic-ray protons.  In Section~\ref{FluxSection}, we describe four models used to characterize the neutrino fluxes originating at the source and arriving at Earth.  Section~\ref{IntSection} covers our means of keeping account of the variety of interaction channels giving rise to showers and muons.  These are combined in Section~\ref{SpectraSection} to yield event spectra that are then normalized to the IceCube count rate.  These normalizations are ultimately applied to calculating the expected flux of cosmic-ray protons for each model and compared to UHECR data in Section~\ref{CRSection}.

%--------------------------------------------------------------------%
\section{Cosmic Neutrino Fluxes}
\label{FluxSection}
We consider four representative models for the shape and flavor composition of arriving neutrino spectra, with normalizations later obtained by matching the IceCube event rate, to describe processes that generate neutrinos in different types of sources.  One or more spectral breaks will typically be necessary, so we make use of a general smoothly-broken power law form to describe the spectra arising from sources with breaks at $E_1$ and $E_2$,
\begin{equation}
      \frac{dN}{dE}  \!=\!   f_i \!
       \left[\left(\frac{E}{E_1}\right)^{\!\alpha \eta} \!\!+\! \left(\frac{E}{E_1}\right)^{\!\beta \eta}
        \!\!+\! \left(\frac{E_2}{E_1}\right)^{\!\beta \eta}\!\! \left(\frac{E}{E_2}\right)^{\!\gamma \eta} \right]^{1/\eta} \!\!\!,
\label{fit2}
\end{equation}
with $\alpha$, $\beta$, and $\gamma$ the slopes, $\eta \!=\! -2$ giving smooth breaks (approximating variation between individual sources), and $f_i$ absorbing a factor of $E_1^\alpha$.  A break might arise due to an intrinsic cutoff in the accelerated proton spectrum or in other ways described below.

We obtain the neutrino fluxes at Earth, $\varphi_\nu(E_\nu)$, by integrating each source spectrum up to $z_{\rm max} \!=\! 8$ as
\begin{equation}
  \varphi_\nu(E_\nu) = \frac{c}{4 \pi } \int_0^{z_{max}} \frac{dN_{\nu}}{dE_{\nu}^\prime}  \frac{dE_{\nu}^\prime}{dE_{\nu}}\, \frac{\mathcal{W}(z)}{dz/dt} \,dz \,,
\label{f1}
\end{equation}
where ${dz}/{dt} \!=\! H_0\, (1 \!+\! z) [\Omega_m (1 \!+\! z)^3 \!+\! \Omega_\Lambda ]^{1/2}$, ($\Omega_m \!=\! 0.3$, $\Omega_{\Lambda} \!=\!0.7$, and ${H}_{0} \!=\! 70\,$km/s/Mpc), and $dE_\nu^\prime/dE_\nu \!=\! (1+z)$ accounts for redshift.  We set the rate evolution, $\mathcal{W}(z) \!=\! 1$, which conservatively bounds the required neutrino emissivity, and discuss alternatives later.

Model~{\color{c1} 1} is motivated by the AGN model in \cite{Mannheim:1998wp} for an $E^{-2}$ accelerated proton spectrum with an $E^{-2}$ target photon background, producing neutrons and pions via $p\, \gamma\! \rightarrow \! N\,\pi$.  Throughout, we assume $E_\nu \! \sim \! 1/20 \,E_n \! \sim \! 10^{-1.3} \,E_n$, with pions and muons yielding identical average neutrino energies (and are not reaccelerated \cite{Klein:2012ug}).

The initial slope of the neutrino spectrum is $\alpha \!=\! -1$, steepening to $\beta \!=\! -2$ at $E_1 \!=\! 10^{6.7}\,$GeV due to the photoproduction opacity growing to $\gtrsim\,$1 at $E_n \!=\! 10^8\,$GeV.  We assume here that equal numbers of $\pi^+$ and $\pi^-$ are produced, appropriate for interactions well above the photoproduction threshold, and that $E_2 \!>\! 10^8\,$GeV.  The decays $\pi^+ \!\rightarrow\! \mu^+ \nu_\mu$, $\mu^+ \!\rightarrow\! e^+ \bar{\nu}_\mu \nu_e$ and $\pi^- \!\rightarrow\! \mu^- \bar{\nu}_\mu$, $\mu^- \!\rightarrow\! e^- \nu_\mu \bar{\nu}_e$, initially give $\nu_e$:$\nu_\mu$:$\nu_\tau$$\,=\,$1:2:0 and $\bar{\nu}_e$:$\bar{\nu}_\mu$:$\bar{\nu}_\tau$$\,=\,$1:2:0.  In all models, we neglect the $\bar{\nu}_e$ flux from $n \!\rightarrow\! p\,e^- \bar{\nu}_e$, which carries much less energy and peaks at energies lower by about two orders of magnitude.

Model {\color{c2} 2} considers the limit where only $\pi^+$ are produced, as with predominantly near-threshold photoproduction (i.e., of $\Delta^+$).  It also incorporates the possibility that the magnetic fields within the sources are strong enough to cause pions and muons to lose appreciable energy due to synchrotron radiation prior to decaying \cite{Rachen:1998fd,Winter:2012xq}.  For each particle, the spectrum breaks where the energy-dependent loss rate exceeds the decay rate.  We consider each separately.  For pions, we assume a softer spectrum with $\alpha \!=\! -1.9$, with a cooling break at $E_1 \!=\! 10^{6.7}\,$GeV to $\beta \!=\! -3.9$ corresponding to $B\!\sim\,$5~kG (and that $E_2 \!>\! 10^8\,$GeV) with $\nu_e$:$\nu_\mu$:$\nu_\tau$$\,=\,$0:1:0 and no $\bar{\nu}$.

Cooling is more severe for muons due to their longer lifetime.  The flux from muon decay has the same shape, but breaks at the lower energy of $E_1 \!=\! 10^{5.5}\,$GeV with $\nu_e$:$\nu_\mu$:$\nu_\tau$$\,=\,$1:0:0 and $\bar{\nu}_e$:$\bar{\nu}_\mu$:$\bar{\nu}_\tau$$\,=\,$0:1:0.  Thus, at high energies in Model {\color{c2} 2} there are practically no antineutrinos.

Model~{\color{c3} 3} takes $\alpha \!=\! -1$ with a low break to $\beta \!=\! -2$ at $E_1 \!=\! 10^4\,$GeV.  A break to $\gamma \!=\! -4$ is put in at $E_2 \!=\! 10^{6.5}\,$GeV (similar to the spectrum discussed by IceCube in \cite{Klein:2013}).  Equal $\pi^+$ and $\pi^-$ numbers are assumed, so this model can also approximate fluxes from $p\,p$ scattering.

Model {\color{c4} 4} also takes $\alpha \!=\! -1$ and $E_1 \!=\! 10^4\,$GeV, only with a steeper $\beta \!=\! -2.2$ that extends to $E_2 \!=\! 10^{8.2}\,$GeV after which $\gamma \!=\! -4$.  This model shows the effects of producing only $\pi^+$ without synchrotron cooling.

Fig.~\ref{fluxes} displays the shapes of the total arriving fluxes of $\nu \!+\! \bar{\nu}$ for all four models (using normalizations obtained in Section~\ref{SpectraSection}).  For Models {\color{c1} 1}, {\color{c3} 3}, and {\color{c4} 4}, the flux of a particular neutrino (or $\bar{\nu}$) flavor $i$ can be found by simply dividing each model curve by $N_{\nu+\bar{\nu}}$ then multiplying by the corresponding $N_i$ value in parentheses in Table~\ref{table:ratios}, which accounts for neutrino oscillation using mixing parameters from \cite{Beringer:1900zz}.  Note that the Model {\color{c2} 2} curve instead results from summing two spectra with either $\pi^+$ or $\mu^+$ flavor ratios (as separately denoted in Table~\ref{table:ratios}).

%%%%%%%%%%%%%%%%%%%%%%%%%%%%%%%%%%%
\begin{figure}[b!]
\includegraphics[width=\columnwidth,clip=true]{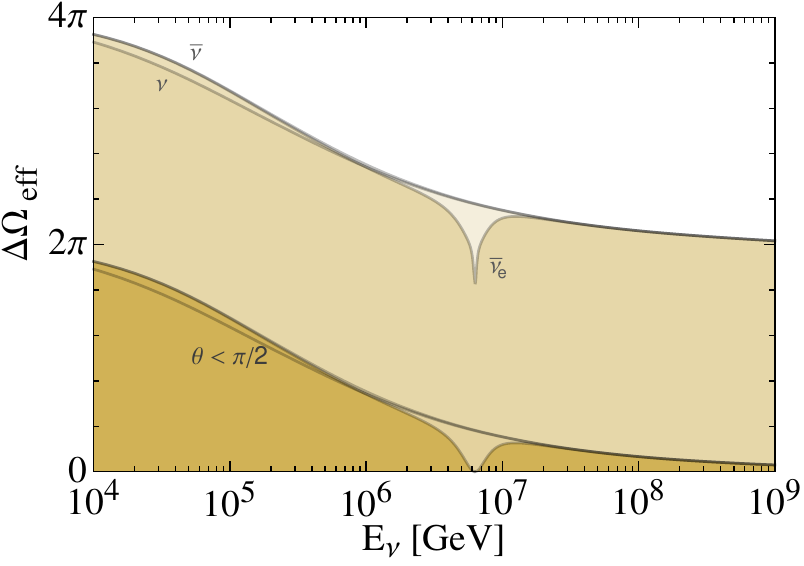}
\caption{Effective solid angle $\Delta\Omega_{\rm eff}$ for $\nu_e$, $\nu_\mu$, and $\nu_\tau$ (``$\nu$''), $\bar{\nu}_\mu$ and $\bar{\nu}_\tau$ (``$\bar{\nu}$''), and $\bar{\nu}_e$ as a function of $E_\nu$.  The upper set of lines average over the full sky, while the bottom set shows the contribution only from angles below the horizon.
\label{coleff}}
\end{figure}
%%%%%%%%%%%%%%%%%%%%%%%%%%%%%%%%%%%

%--------------------------------------------------------------------%
\section{Neutrino Interactions}
\label{IntSection}
The most relevant quantity for interactions in IceCube is the amount of visible radiation produced, which depends upon the interaction channel.  We work in terms of the electromagnetic-equivalent energy, $E_{\rm em}$, defined such that for an electron $E_{\rm em}\!=\!E_e$.  For neutrino-nucleon, $\sigma_{\nu N}$ and $\sigma_{\bar{\nu} N}$,  interactions, we use the total deep-inelastic scattering cross sections from \cite{Gandhi:1998ri} for charged-current (CC) and neutral-current~(NC) scattering and approximate the average inelasticity $\langle y(E_\nu)\rangle$ \cite{Gandhi:1995tf} as 0.25.

%%%%%%%%%%%%%%%%%%%%%%%%%%%%%%%%%%%
\begin{figure*}[t]
\includegraphics[width=0.99\columnwidth,clip=true]{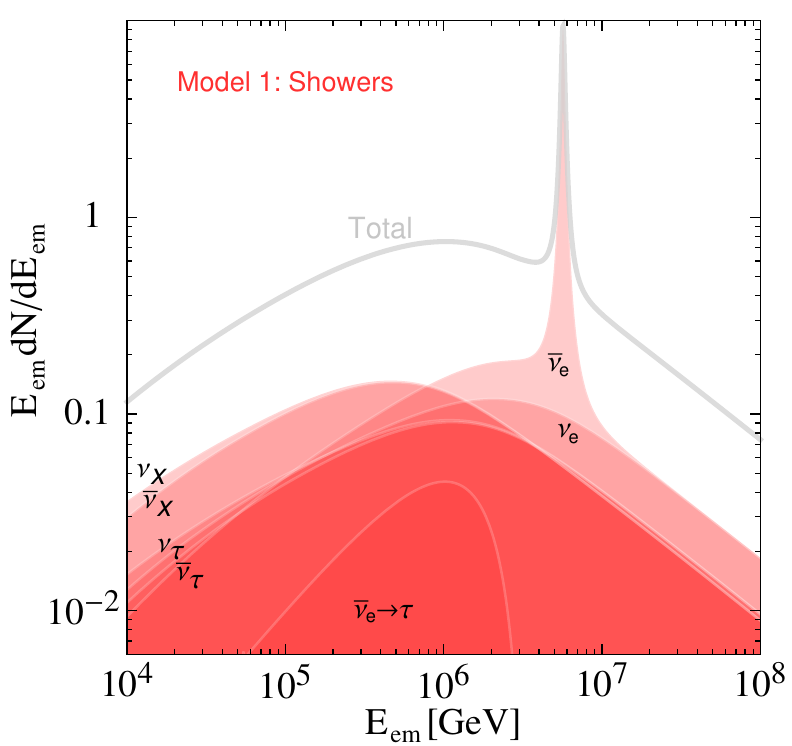}
\includegraphics[width=0.99\columnwidth,clip=true]{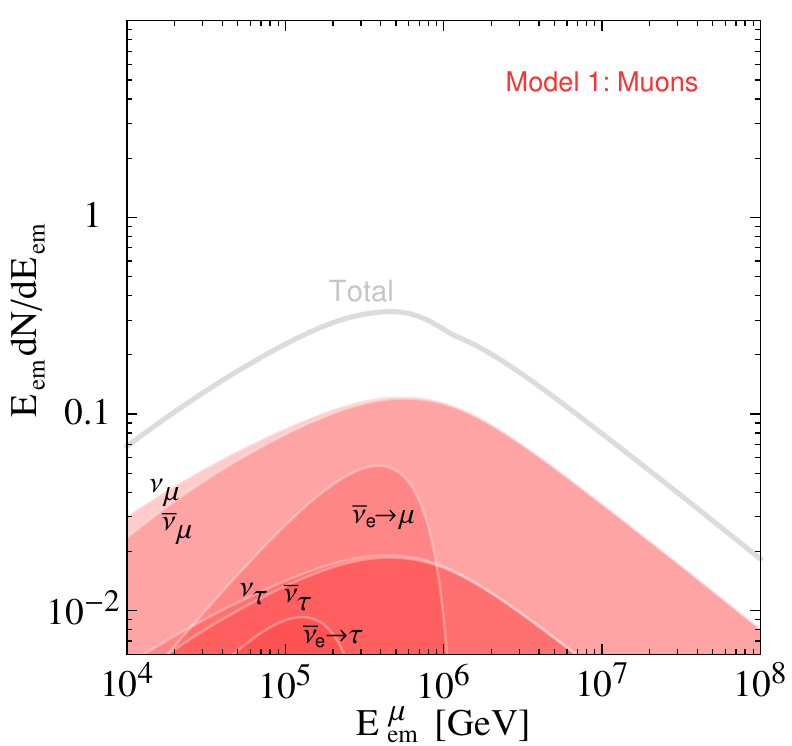}
\vspace{0.0in}
\includegraphics[width=0.99\columnwidth,clip=true]{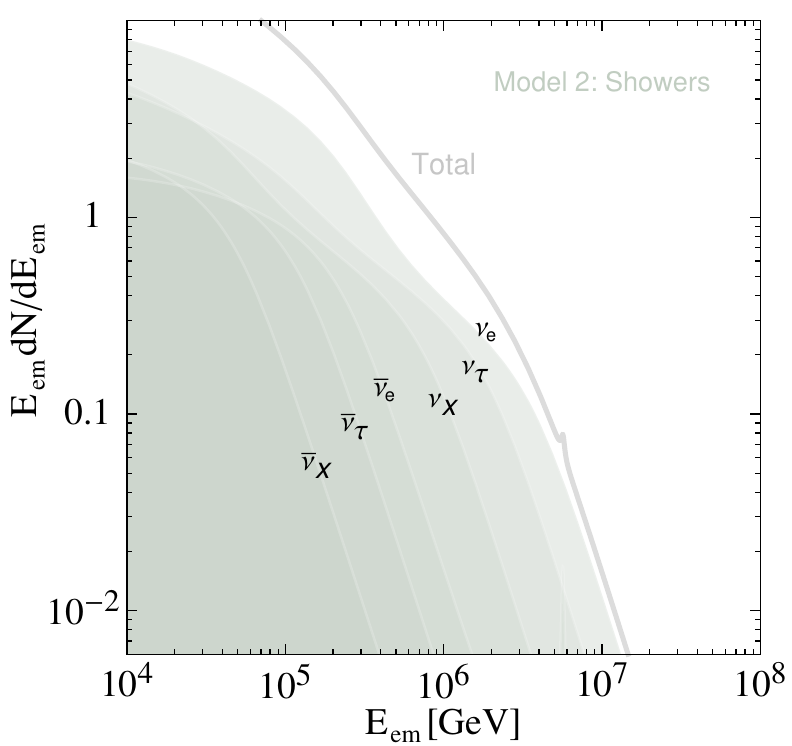}
\includegraphics[width=0.99\columnwidth,clip=true]{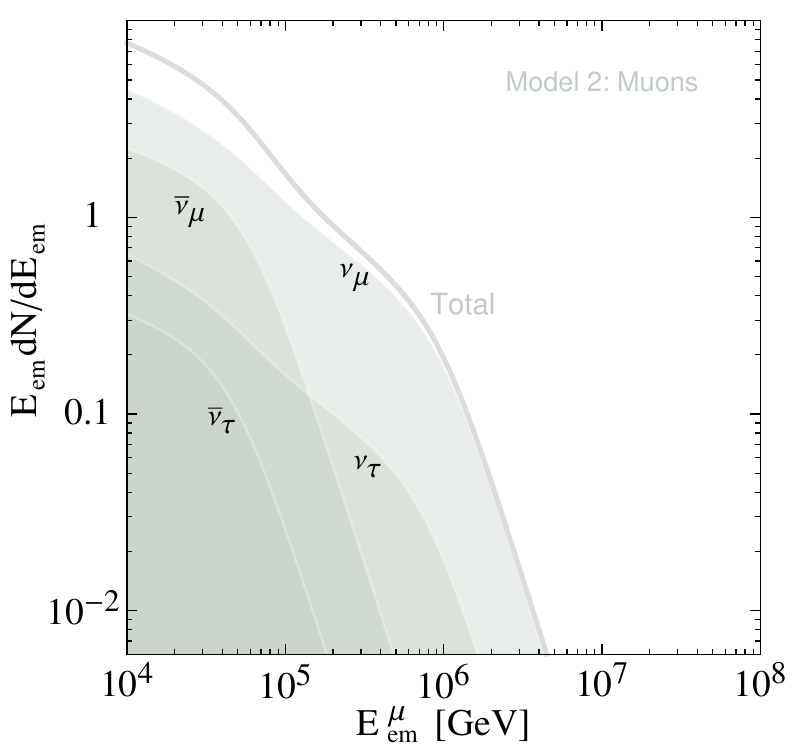}
\caption{Neutrino event spectra for Models {\color{c1} 1} and {\color{c2} 2} versus the electromagnetic-equivalent $E_{\rm em}$ for showers ({\it left panels}) and $E^\mu_{\rm em}$ for contained-vertex muons ({\it right panels}).  Shower components (as labeled) include $\nu_e$, $\bar{\nu}_e$, $\nu_\tau$, and $\bar{\nu}_\tau$ charged current (CC); all flavor ($\nu_X$, $\bar{\nu}_X$) neutral current; and $\bar{\nu}_e e$ Glashow resonance channels yielding $e$ and hadrons (included in $\bar{\nu}_e$ line) and $\tau$.  Muon components include $\nu_\mu$ and $\bar{\nu}_\mu$ CC; $\nu_\tau$ and $\bar{\nu}_\tau$ CC decaying to muons; and $\bar{\nu}_e e$ channels yielding a $\mu$ or $\tau$.
The total neutrino fluxes are shown in Fig.~\ref{fluxes}, normalized to three total $>\,$PeV events in IceCube for Model~{\color{c1} 1} and ten $>\,$100~TeV events for Model~{\color{c2} 2}.  Note that the curves would decrease somewhat below $\sim\! 10^5\,$GeV if energy dependence in $V_{\rm eff}$ \cite{Whitehorn} was incorporated rather than $V_{\rm eff} \!=\! 0.4\,$km$^{-3}$.
\label{show1}}
\end{figure*}
%%%%%%%%%%%%%%%%%%%%%%%%%%%%%%%%%%%

%%%%%%%%%%%%%%%%%%%%%%%%%%%%%%%%%%%
\begin{figure*}[t]
\includegraphics[width=0.99\columnwidth,clip=true]{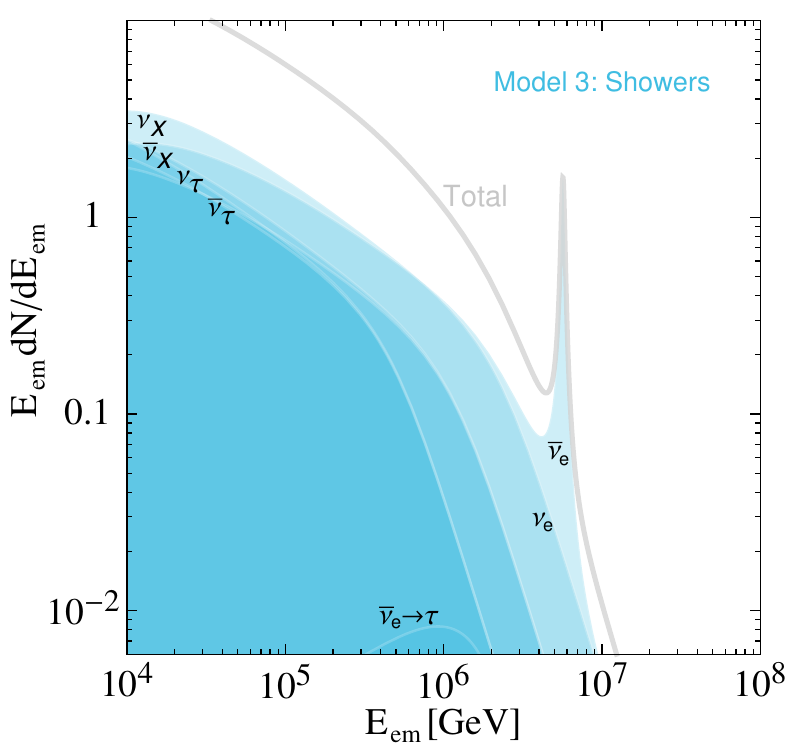}
\includegraphics[width=0.99\columnwidth,clip=true]{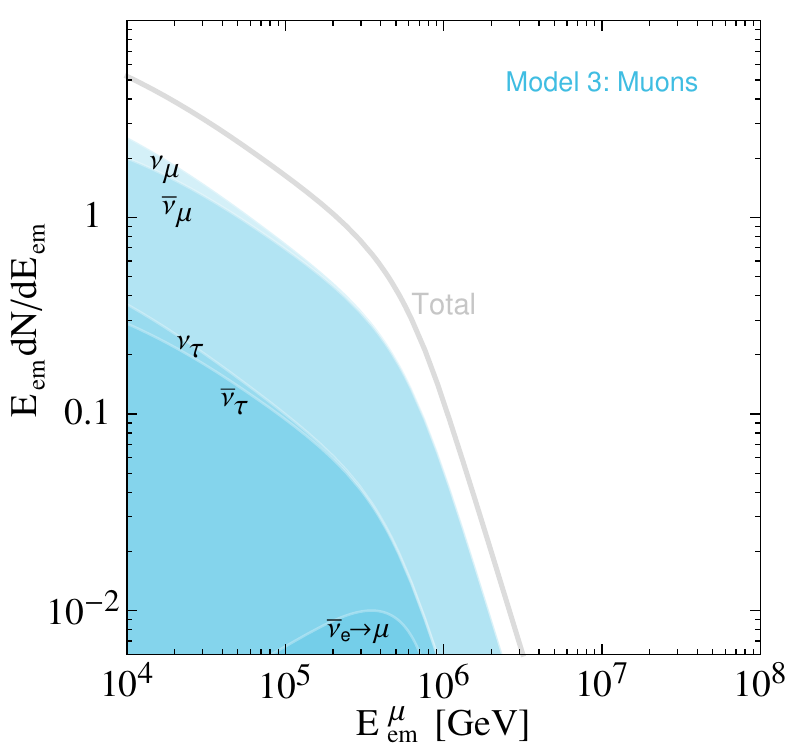}
\vspace{0.0in}
\includegraphics[width=0.99\columnwidth,clip=true]{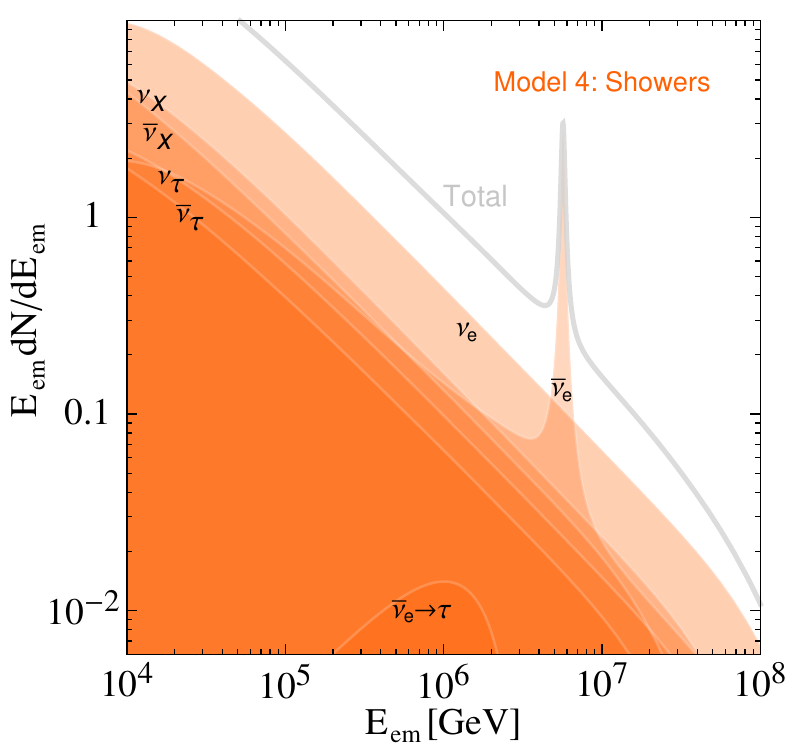}
\includegraphics[width=0.99\columnwidth,clip=true]{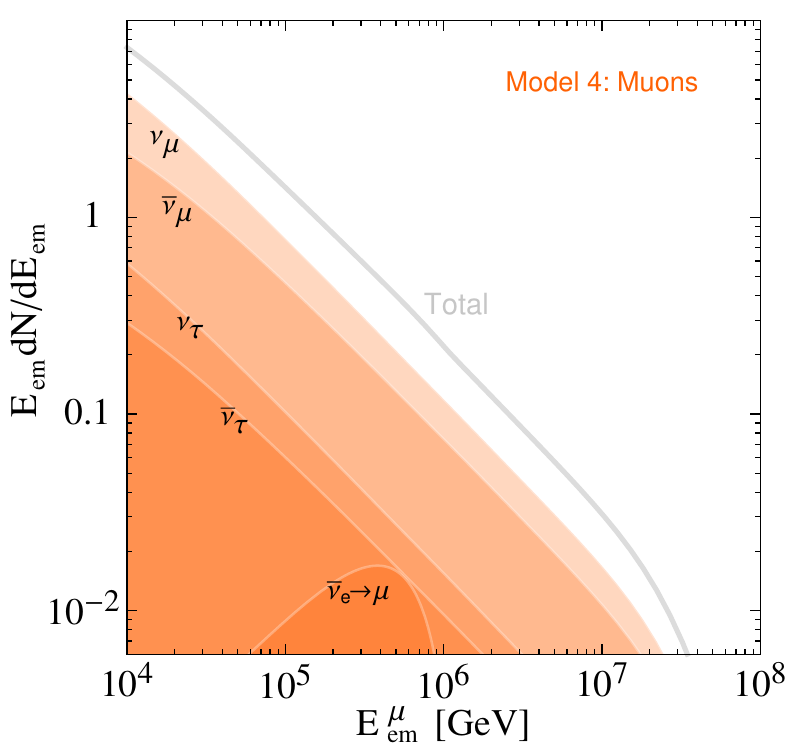}
\caption{Neutrino event spectra for Models {\color{c3} 3} and {\color{c4} 4} versus the electromagnetic-equivalent $E_{\rm em}$ for showers ({\it left panels}) and $E^\mu_{\rm em}$ for contained-vertex muons ({\it right panels}).  Shower components (as labeled) are the same as in Fig.~\ref{show1}.  The total neutrino fluxes from both models are also shown in Fig.~\ref{fluxes}, normalized to ten total $>\,$100~TeV events.
\label{show3}}
\end{figure*}
%%%%%%%%%%%%%%%%%%%%%%%%%%%%%%%%%%%

\vspace{-0.2in}
%--------------------------------------------------------------------%
\subsection{Showers}
\vspace{-0.1in}
In CC $\nu_e$ events ($\nu_e\,N$$\, \rightarrow \,$$e\,X$), we assume that the electron deposits its entire energy, $E_e$$\, = \,$$\langle 1 - y\rangle E_\nu$, into an electromagnetic shower.  The recoiling nucleon receives a fraction $\langle y \rangle E_\nu$ of the energy, resulting in a hadronic shower that yields less light than an equivalent-energy electromagnetic shower by a factor that is a function of energy \cite{Kowalski:2004qc,Aartsen:2013vja} assumed to be $f_{\rm had}\simeq 0.9$.  We add these together to get a total effective visible energy per interaction as $E_{{\rm em,}\,e}$$\, = \,$$\langle 1 - y \rangle E_\nu + f_{\rm had} \langle y \rangle E_\nu$$\, = \,$$q_e\,E_\nu$, with $q_e$$\, \approx \,$$0.95$ (and similarly for $\bar{\nu}_e$).

For NC events ($\nu\,N$$\, \rightarrow \,$$\nu\,X$), $\sigma_{\rm NC}$ is identical for all flavors (although the cross section for $\nu$ is somewhat larger than for $\bar{\nu}$).  The only visible energy in this case is due to the hadronic shower from the recoiling nucleon.  For $E_{\rm em, NC} \!=\! q_{\rm NC}\,E_\nu$, we take $q_{\rm NC}$$\, \approx \,$$0.2$.

The properties of tau neutrino CC events ($\nu_\tau\,N$$\, \rightarrow \,$$\tau\,X$) depend upon the subsequent decay of the $\tau$.  Of the decays, $\sim 17$\% go directly to muons ($\tau\! \rightarrow \!\mu\,\nu_\mu\,\nu_\tau$), with the other channels \cite{Beringer:1900zz} resulting in a shower.  In each, a $\nu_\tau$ leaves with a fraction of the total energy.  The $\tau\! \rightarrow \!e\,\nu_e\,\nu_\tau$ channel yields an electromagnetic cascade with an additional outgoing $\nu$.  Other decays involve multiple mesons, which result in a hadronic shower.  This heterogeneity leads to a broad range of light output.  We assume all such decays to give cascades intermediate between CC $\nu_e$ and NC events, with $E_{{\rm em,}\,\tau}$$\, = \,$$q_\tau \,E_\nu$, where $q_\tau$$\,\approx \,$$0.5$.  While the $\tau$ can potentially distinctively travel a measurable distance between the initial hadronic cascade and its decay ($\sim\,$50~m at 1~PeV) \cite{Learned:1994wg}, we take the initial and decay bangs to be indistinguishable.

While scattering on electron targets is usually negligible, a unique spectral feature can arise from the interactions of $\bar{\nu}_e$.  This is due to the Glashow resonance, $\bar{\nu}_e e$$\,\rightarrow \,$$W^-$$\rightarrow \,$$X$, which results in a sharply enhanced $\sigma_{\bar{\nu} e}$ near $E_{\bar \nu_e}$$\,\approx\,$6.3$\,$PeV.  $W$ decay channels yielding quarks are purely hadronic ($q_{G,q}$$\,\approx \,$$0.9$).  The $e$/$\tau$ channels result in a neutrino that carries away most of the energy ($\langle y \rangle$$\, \approx \,$$0.25$ \cite{Gandhi:1995tf}).

The spectrum of events from each channel is given in terms of the electromagnetic-equivalent energy as~\cite{Gaisser:1990vg,Kistler:2006hp,Beacom:2007yu}
\begin{equation}
  \frac{dN_{\rm sh}}{dE_{\rm em}} = N_A\,\rho\,T\, V_{\rm eff}\, \Delta\Omega_{\rm eff}\, \sigma(E_\nu)
  \, \varphi_\nu(E_\nu)/q \,,
  \label{showeq}
\end{equation}
where $N_A\,\rho$ is the molar density of ice.  The effective volume of IceCube for showers is $V_{\rm eff}\!\approx\!0.4\,$km$^3$ after reaching full efficiency at $E_{\rm em}\!\gtrsim\,100\,$TeV \cite{Whitehorn} and we use $T\!=\!988$~days \cite{Klein:2013}.  For $\bar{\nu}_e e$ events, it is important to have a better handle on the spectral features from the Glashow resonance, so we instead use the full differential cross section \cite{Gandhi:1995tf}.  Note that the relevant target density in this channel is that of electrons, which is lower by a factor of 10/18.  Our estimates using the $d\sigma/dy$ distributions in \cite{Gandhi:1995tf} for other channels agree at the $\sim\,$10\% level.

The scattering of neutrinos within the Earth lead to an effective solid angle $\Delta\Omega_{\rm eff}\!=\!4\pi\,\langle e^{-\tau_\oplus} \rangle \!<\! 4\pi$, with $\tau_\oplus=N_A\, \lambda_\oplus\,\sigma_{\rm tot}(E_\nu)$.  For $\nu_e$, $\nu_\mu$, and $\nu_\tau$, $\sigma_{\rm tot} \!=\! \sigma_{\nu N}$.  $\sigma_{\rm tot} \!=\! \sigma_{\bar{\nu} N}$ for $\bar{\nu}_\mu$ and $\bar{\nu}_\tau$, although for $\bar{\nu}_e$ the Glashow resonance contribution is also included.  The variation as a function of $E_\nu$ for each neutrino species, using the column depth from the Preliminary Reference Earth Model \cite{Dziewonski:1981xy} and averaging over the full sky, is shown in Fig.~\ref{coleff}.  While NC interactions result in loss of energy without removing the actual neutrino from the beam (similarly with $\nu_\tau$), we do not attempt to account for this here since the typical neutrino spectrum is declining.

\vspace{-0.2in}
%--------------------------------------------------------------------%
\subsection{Muons}
\vspace{-0.1in}
Muon neutrino CC scatterings ($\nu_\mu\,N \rightarrow \mu\,X$) also produce an initial hadronic shower, with a defining characteristic of such events being a resulting outwardly-directed muon track.   The $\sim\!17$\% of tau decays that result in a muon possess similar characteristics, although the muon carries less energy than that resulting from a CC $\nu_\mu$ event with a shower of equivalent energy.  We take $E_\mu^{\tau\rightarrow \mu} \!=\! \langle 1-y \rangle/3 \, E_\nu \!=\! q_{\tau \mu} \,E_\nu$, with $q_{\tau \mu}$$\,\approx \,$$0.25$.

While one can calculate the expected energy spectrum of muons that are produced, a more relevant quantity to compare to measurable quantities in IceCube is the deposited energy, which will include contributions from both the birth shower and muon energy losses.  To approximate the energy lost by the muon during propagation within the detector, we first consider an average continuous energy loss of \cite{Lipari:1991ut,Beringer:1900zz}
\begin{equation}
	\frac{dE}{dX}=-\alpha_\mu - \beta_\mu E_\mu\,.
\end{equation}
Integrating over a given range $R_\mu$ in ice, where $\alpha_\mu\!=\!2.0\!\times\!10^{-6}\,$TeV~cm$^2$~g$^{-1}$ and $\beta_\mu\!=\!4.2\!\times\!10^{-6}\,$cm$^2$~g$^{-1}$, and adding the hadronic shower, gives a total EM equivalent energy deposited of
\begin{eqnarray}
  E_{\rm em}^\mu &=& f_{\rm had} \langle y \rangle E_\nu + E_\mu^i - E_\mu^f \\
                             &\approx& E_\nu - \{ {\rm Exp}[{\rm ln}(\alpha_\mu + \beta_\mu E_\mu^i) \!-\! \beta R_\mu] \!-\! \alpha_\mu \}/\beta_\mu \nonumber
     \,,
  \label{mushoweq}
\end{eqnarray}
where $E_\mu^i \!=\! \langle 1 - y \rangle E_\nu$ is the initial energy of the muon and $E_\mu^f$ is its energy as it exits the detector.  For $E_\nu \!\gtrsim\! 10$~TeV, losses due to ionization (the $\alpha_\mu$ term) are subdominant (these generally do not lead to observable energy deposition anyway in IceCube).  Using a typical range of $R_\mu\!\sim\,$0.5~km and treating losses due to pair production, bremsstrahlung, and photohadronic interactions as purely electromagnetic would lead to $E_{\rm em}^\mu \!\approx\! 0.4\,E_\nu$.  This probably overestimates the light yield, and we find $E_{\rm em}^\mu \!\approx\! 0.25\,E_\nu$ better agrees with IceCube.

Muons arising from a Glashow resonance event lack an initial hadronic shower.  For this case, we take $E_{\rm em}^\mu \!\approx\! 0.2\,E_\mu^i$.  The proportionality of the energy loss rate to the muon energy, the high average initial muon energy of $\sim\,$1.5~PeV, and the stochasticity of radiative losses could conspire to result in a large energy deposition near the start of the track, although more sophisticated techniques may have adequate discriminating power \cite{Aartsen:2013vja}.  We also assume that hadronic cascades will not themselves yield an energetic muon track.

%%%%%%%%%%%%%%%%%%%%%%%%%%%%%%%%%%%
\begin{figure}[b!]
\includegraphics[width=\columnwidth,clip=true]{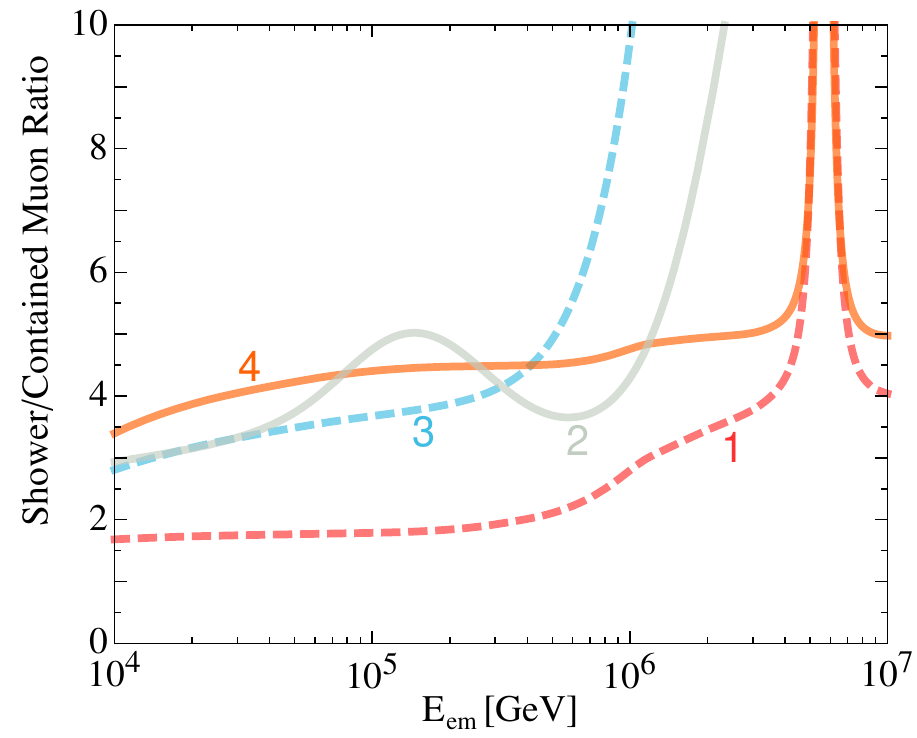}
\caption{Ratio of neutrino shower to contained-vertex muon events versus deposited energy for Models {\color{c1} 1}, {\color{c2} 2}, {\color{c3} 3}, and {\color{c4} 4}.
\label{flavar}}
\end{figure}
%%%%%%%%%%%%%%%%%%%%%%%%%%%%%%%%%%%

%--------------------------------------------------------------------%
\section{Neutrino Event Spectra}
\label{SpectraSection}
Using the above method along with our four neutrino spectral models, we calculate the event spectra of each interaction channel in IceCube.  We consider only events with $>\,$100~TeV EM-equivalent energy deposition in the detector, where the effective volume is nearly constant and backgrounds are low.  These total 12 IceCube events, one of which has an upgoing muon track ($E_{\rm em} \!\sim\,$250~TeV), with three shower events at $>\,$1~PeV \cite{Klein:2013}.  The estimated background is $\sim\! 1 \!-\! 3$ events \cite{Klein:2013}.  Saturating a previous 90\% CL upper limit on the prompt atmospheric neutrino flux from charmed mesons \cite{Aartsen:2013eka} would only yield $\sim\!3.5$ events, although the best model fit obtained by IceCube contains no prompt component \cite{Klein:2013}.  At $>\,$1~PeV, no backgrounds yield $\gg\!0.01$ events \cite{Klein:2013}.

Figs.~\ref{show1} and~\ref{show3} show the resulting spectrum for each $\nu N$, $\bar{\nu} N$, and $\bar{\nu}_e e$ shower and muon channel in the detector.  Model~{\color{c1} 1} peaks at $\sim\,$1~PeV by design, so we obtained the normalization by integrating the total shower and muon curves above 1~PeV, summing them, and equating to three events, implicitly assuming that the lower energy events arise from another distinct flux (although choosing a lower cutoff would not change much).  This results in $f_1 \! \approx \! 1.5 \!\times\! 10^{-48}\,$GeV$^{-1}$cm$^{-3}$s$^{-1}$ when using Eq.~(\ref{fit2}).

For the other models, we integrate and sum both totals above 100~TeV and equate to ten counts, not attempting a detailed spectral fit.  We break Model~{\color{c2} 2} into $\pi$ and $\mu$ decay components, with $f_{2,\pi} \! \approx \! 1.3 \!\times\! 10^{-48}\,$GeV$^{-1}$cm$^{-3}$s$^{-1}$ and $f_{2,\mu} \! \approx \! 4.8 \!\times\! 10^{-46}\,$GeV$^{-1}$cm$^{-3}$s$^{-1}$.  In Model~{\color{c3} 3}, $f_3 \! \approx \! 4.4 \!\times\! 10^{-43}\,$GeV$^{-1}$cm$^{-3}$s$^{-1}$.  For Model~{\color{c4} 4}, $f_4 \! \approx \! 8.6 \!\times\! 10^{-43}\,$GeV$^{-1}$cm$^{-3}$s$^{-1}$.  Note the differing overall scaling related to varying the $E_1^\alpha$ values in Eq.~(\ref{fit2}).

IceCube's fit assuming an $E_\nu^{-2}$ spectrum up 3~PeV, a 1:1:1 flavor ratio, and equal numbers of $\nu$ and $\bar{\nu}$ yielded a per-flavor ($\nu \!+\!\bar{\nu}$) flux of $E_\nu^{-2} \varphi_\nu \!=\! 0.95 \!\pm\! 0.3 \!\times\! 10^{-8}\,$GeV~cm$^{-2}\,$s$^{-1}\,$sr$^{-1}$ \cite{Klein:2013}.  Multiplying by three, we see that the normalization agrees well with our Model~{\color{c3} 3}.  Using an energy-dependent $V_{\rm eff}$ such as shown in \cite{Whitehorn} would result in relatively fewer expected events at $E_{\rm em} \!\lesssim\! 200$~TeV and raise our required normalization by $\sim10\,$\%.

Examining the importance of the Glashow resonance (see also \cite{Beacom:2003nh,Anchordoqui:2004eb,Barger:2012mz,Bhattacharya:2012fh,Barger:2014iua}), we see that the distinct peak at $\sim\,$6~PeV due to the quark decay channel is most prominent in Model~{\color{c1} 1}, while completely absent in Model~{\color{c2} 2}.  The $d\sigma/dy$ distribution significantly broadens the other decay channels.  The $W$ width to hadrons is a factor of $\sim\,$3 larger than the sum of $e$/$\tau$, so 6~PeV showers are the most likely.  The larger attenuation of the $\bar{\nu}_e$ flux within the Earth (as seen in Fig.~\ref{coleff}) results in a minor suppression.  In Model~{\color{c4} 4}, $\bar{\nu}_e$ only arise from oscillations and resonance events are thus relatively lower by a factor of $\sim\!2\!-\!3$.

Fig.~\ref{fluxes} also shows the IceCube flux model based on a piecewise parametrization with approximate 68\% confidence ranges (points and upper limits) from \cite{Klein:2013}.  Note that this also assumes a 1:1:1 flavor ratio and equal numbers of $\nu$ and $\bar{\nu}$, so cannot be directly compared with Model~{\color{c2} 2} or Model~{\color{c4} 4}, since, e.g., their relatively lower $\bar{\nu}_e$ fluxes should ease the strong upper limit at $\sim\!5 \!-\!6$~PeV due to the lack of obvious Glashow resonance events.

In Fig.~\ref{flavar}, we show the ratio of shower-type events to contained muons as a function of energy deposited.  We see that Models {\color{c3} 3} and {\color{c4} 4} both are consistent with the single $\sim\,$250~TeV IceCube muon event observed in the $E_{\rm em}^\mu \!>\! 100\,$TeV range \cite{Klein:2013} thus far.

%%%%%%%%%%%%%%%%%%%%%%%%%%%%%%%%%%%
\begin{figure}[b!]
\includegraphics[width=0.99\columnwidth,clip=true]{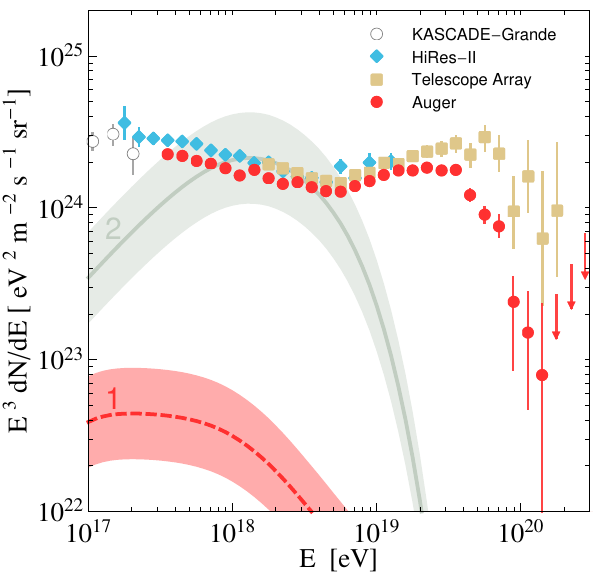}
\caption{The ultrahigh-energy cosmic-ray spectrum.  Shown are the proton fluxes associated with neutrino Models {\color{c1} 1} ({\it dashed}) and {\color{c2} 2} ({\it solid}).  Normalizations are obtained from IceCube neutrino data (as described in the text) with 50\% uncertainty bands.  These are compared to KASCADE-Grande~\cite{Apel:2013dga}, HiRes-II \cite{Abbasi:2007sv}, Auger \cite{Abreu:2011pj}, and Telescope Array \cite{AbuZayyad:2012ru} data.
\label{UHECR}}
\end{figure}
%%%%%%%%%%%%%%%%%%%%%%%%%%%%%%%%%%%

%--------------------------------------------------------------------%
\section{The cosmic-ray spectrum}
\label{CRSection}
With a few simplifying assumptions, we relate our neutrino spectra to a proton flux with a normalization that is thus fixed by IceCube.  Since the highest deposited energy seen is presently $\sim\,$2~PeV \cite{Klein:2013}, if the neutrino came from $z \!\sim\! 1$, the source proton must have had $E_p \!\gtrsim\! 10^{17}\,$eV, so we are indeed already in the UHECR ballpark.

As stated earlier, we assume that $E_n\!\sim\!20\, E_\nu$, with the number of neutrinos associated with each neutron decaying to a proton with $E_p\!\approx\!E_n$ varying between models.  In Model~{\color{c1} 1}, we have assumed that one pair of $\pi^\pm$ is produced, so that six total neutrinos result from the $\pi^\pm$$\mu^\pm$ decays.  The conversion to a proton flux is easiest to interpret at energies below the first break.  Here, we simply write
\begin{equation}
  E_p \frac{dN}{dE_p}  = \frac{1}{6} E_\nu \frac{dN}{dE_\nu}
\end{equation}
with neutron decay giving a source proton spectrum in the form of Eq.~(\ref{fit2}) with $\alpha \!=\! -1$.  Above $E_{1p} \!\approx\! 10^{8}\,$GeV, the spectrum steepens by 1, as is reflected in the corresponding neutrino spectrum, however neutron interactions cause another steepening at a slightly higher energy \cite{Mannheim:1998wp}, so that we simply assume $\beta \!=\! -3$.  We insert a high-energy cutoff by choosing $\gamma \!=\! -4$ above $E_{2p} \!\approx\! 10^{9.5}\,$GeV.  Integrating this spectrum over $10^{13} \!<\! E_p \!<\! 10^{20}\,$eV, we find an emissivity $\mathcal{E}_{p{\color{c1} 1}} \sim 3.3 \times 10^{44}~\rm{erg~Mpc}^{-3}\,\rm{yr}^{-1}$, which isn't greatly affected by the higher energy break.

The proton flux associated with Model~{\color{c2} 2} is taken to have the same spectral slope as the pion component, $\alpha \!=\! -1.9$, but allowed to continue to higher energies since, while the neutrino flux is steepened by pion synchrotron losses, the neutron decay length $\gamma_n \,c \,\tau_n \!\sim\! 10$~kpc for $E_n \!\sim\! 10^{18}\,$eV is supposed sufficient to allow escape from the magnetized loss region.  In this case, it is easiest to think in terms of the pionic spectrum (with the muon component modulating the overall normalization) and again begin at low energies.  Now each proton is associated with a single neutrino species, rather than six as in Model~{\color{c1} 1}, giving a factor of six relative increase.  We include only an exponential cutoff at $E_p \!\sim\! 10^{18.5}\,$eV, where synchrotron losses by protons in the magnetic field needed for this case would be relevant (see \cite{Aharonian:2000pv,Winter:2013cla}).  In total $\mathcal{E}_{p{\color{c2} 2}} \!\sim\! 8.9 \!\times\! 10^{45}~\rm{erg~Mpc}^{-3}\,\rm{yr}^{-1}$.

The proton fluxes for Models {\color{c3} 3} and {\color{c4} 4} are more straightforward.  Model {\color{c3} 3} takes $\alpha \!=\! -1$, $\beta \!=\! -2$, and $\gamma \!=\! -4$, with $E_{1p} \!\approx\! 10^{5.3}\,$GeV and $E_{2p} \!\approx\! 10^{7.8}\,$GeV, with six neutrinos per proton.  Here $\mathcal{E}_{p{\color{c3} 3}} \!\sim\! 1.4 \!\times\! 10^{45}~\rm{erg~Mpc}^{-3}\,\rm{yr}^{-1}$.  Model~{\color{c4} 4} uses $\alpha \!=\! -1$, $\beta \!=\! -2.2$, and $\gamma \!=\! -4$, with $E_{1p} \!\approx\! 10^{5.3}\,$GeV and $E_{2p} \!\approx\! 10^{9.5}\,$GeV.  There are three neutrinos per proton and $\mathcal{E}_{p{\color{c4} 4}} \!\sim\! 4.2 \!\times\! 10^{45}~\rm{erg~Mpc}^{-3}\,\rm{yr}^{-1}$.

%%%%%%%%%%%%%%%%%%%%%%%%%%%%%%%%%%%
\begin{figure}[t!]
\includegraphics[width=0.99\columnwidth,clip=true]{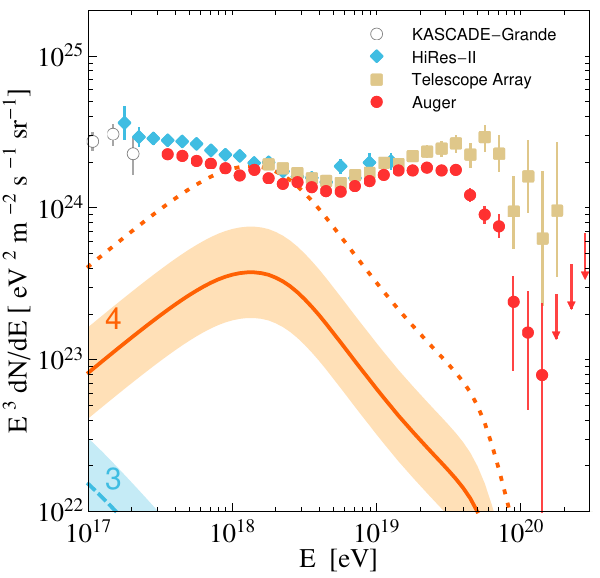}
\caption{Proton fluxes associated with neutrino Models {\color{c3} 3} ({\it dashed}) and {\color{c4} 4} ({\it solid}), in the same fashion as described in Fig.~\ref{UHECR} and compared to the same UHECR data.  Boosting the flux of Model~{\color{c4} 4} by a factor of 5 results in the {\it dotted} curve.
\label{UHECR3}}
\end{figure}
%%%%%%%%%%%%%%%%%%%%%%%%%%%%%%%%%%%

Calculating the expected proton fluxes at Earth from these spectra in order to compare with cosmic-ray data (shown in Figs.~\ref{UHECR} and \ref{UHECR3}) is somewhat more involved than for neutrinos from Eq.~(\ref{f1}), since more means of losing energy are open to protons.  Three types of energy loss are accounted for in propagation.  At energies where pion photoproduction on the CMB occurs ($ \gtrsim \,$10$^{19.5}\,$eV), $p \, \gamma \rightarrow N \, \pi$ is the dominant loss channel~\cite{Mucke:1998mk}.  For $E_p \gtrsim 10^{18}$~eV, and below the CMB photopion threshold, resonant pair production on background photons, $p \, \gamma \rightarrow p\, e^+e^-$, dominates~\cite{Blumenthal:1970nn}.  This process has a large cross section, although each interaction removes only a small amount of energy.  Finally, there is an adiabatic redshift loss term. 

These can be combined via characteristic loss times \cite{Berezinsky:2002nc} as $\tau_T^{-1}(E_p,z)$$\, = \,$$\tau_\pi^{-1}(E_p,z) + \tau_{\rm pair}^{-1}(E_p,z) + \tau_a^{-1}(z)$, giving an energy loss rate of $d\ln{E_p}/dt$$\, = \,$$\tau_T^{-1}(E_p,z)$.  We can thus relate the injection energy at redshift $z$, $E^\prime_p=E^\prime_p(E_p,z)$, to the detected energy, $E_p$, with
\begin{equation}
  \frac{1}{E_p}\frac{dE_p}{dz}=\frac{1}{\tau_T(E_p,z)}\frac{1}{dz/dt} \,.
\label{loss}
\end{equation}
The constraints imposed by this relation can be seen in Fig.~3 of Ref.~\cite{Yuksel:2006qb}.  Using the same $\mathcal{W}(z)$ as above, we calculate the spectrum of the arriving proton flux as
\begin{equation}
  \varphi_p(E_p)  =  \frac{c}{4 \pi} \int_0^{z_{max}} \frac{dN_{p}}{dE_{p}^\prime} \frac{\partial E_p^\prime}{\partial E_p} \frac{\mathcal{W}(z)} {dz/dt} dz \,,
\end{equation}
with ${\partial E_p^\prime(E,z)}/{\partial E_p}$ calculated numerically from Eq.~(\ref{loss}).

In Figs.~\ref{UHECR} and \ref{UHECR3}, we present the expected cosmic-ray proton spectra that result from our four models normalized to IceCube, which we discuss below, with 50\% uncertainty bands comparable to the Poisson uncertainties on the neutrino counts \cite{Gehrels:1986mj}.  We have also checked, using CRPropa \cite{Kampert:2012fi}, that the cosmogenic gamma-ray fluxes are safely below the Fermi isotropic background \cite{Abdo:2010nz} and associated constraints \cite{Berezinsky:2010xa,Gelmini:2011kg,Ahlers:2010fw,Decerprit:2011qe,Murase:2012df}.  For $E_p \!\lesssim\! 10^{17}\,$eV, these can be interpreted as upper limits since $\gamma_n \,c \,\tau_n \!\lesssim\! 1$~kpc does not necessarily imply freedom from the host galaxy.

%--------------------------------------------------------------------%
\section{Discussion and conclusions}
If cosmic rays in the $10^{18}\,$eV range are truly extragalactic protons, the steepness of the UHECR spectrum ($\sim\! E^{-3}$) implies that their cosmic energy density is much larger than those at the highest measured energies.  If their escapes from acceleration regions were facilitated by photoproduction of neutrons, then there must also be a substantial flux of neutrinos from pion decays.  The simplified models that we have examined capture the general flavor of the neutrino flux needed to explain the IceCube PeV events and serve as a basis for comparison with cosmic-ray protons.

Fig.~\ref{UHECR} displays the cosmic-ray proton flux associated with the AGN-motivated neutrino fluxes.  We see that Model {\color{c1} 1} is significantly lower than the data (and the $\nu$/$p$ ratio could be larger \cite{Mannheim:1998wp,Ahlers:2009rf} pushing this even lower).  This model also does not well reproduce the broadband IceCube neutrino counts or the shower/muon ratio.  Attempts to adjust the model parameters, e.g., choosing an initial cutoff at a lower energy, tend to decrease the $10^{18}\,$eV proton flux.

Model {\color{c2} 2} appears to be a better match to measurements in the $10^{18}\,$eV range.  The restrictions on this scenario result in this larger flux when the neutrinos are normalized to IceCube.  These include a need for near-threshold photoproduction to only produce $\pi^+$. The low-energy spectrum is also softer than Model {\color{c1} 1}, requiring softer proton and/or harder photon spectra in the source.  A magnetic field of $B \!\sim\!1\!-\!10$~kG cools pions and muons prior to decay, although such fields may occur near jet launching regions \cite{O'Sullivan:2009ua} or the inner accretion disk, where AGN core models operate \cite{Stecker:1991vm,Stecker:2005hn,Stecker:2013fxa}.  Removing the no-$\bar \nu$ requirement would reduce the proton flux by $\gtrsim\,$3, since the Glashow resonance channel reappears, although this can be somewhat compensated for by varying the spectrum.  If the emission is beamed, so that $\Gamma \!>\! 1$, the characteristic breaks would move to lower energies in the frame of the accelerator, requiring a larger $B$ (as perhaps in GRBs \cite{Baerwald:2013pu,Petropoulou:2014lja}).  Spectra in hadronic blazar models thus tend to be hard (e.g., \cite{Muecke:2002bi,Atoyan:2002gu,Dimitrakoudis:2012uf,Dimitrakoudis:2013tpa}).  The shower/muon ratio leads to unique features, with a lack of Glashow resonance events.

Fig.~\ref{UHECR3} shows the fluxes from the neutrino models with parameters chosen to more directly fit the wider IceCube data.  Model {\color{c3} 3} was chosen to have an intrinsic break in the spectrum in order to suppress the neutrino event rate above $\sim\!2$~PeV.  Obviously, if this is directly associated with a decline in the proton flux, there is no way to produce appreciable $10^{18}\,$eV cosmic rays.

Model {\color{c4} 4} appears more promising.  The steeper spectrum leads to relatively fewer high-energy neutrino events, with the lower $\bar{\nu}_e$ flux making the Glashow resonance less prominent, although it is allowed to continue to higher energies.  Even in this case, though, the proton flux does not match up with the UHECR data.

Interpreting our neutrino fluxes as the result of the photopion process, we see that the resulting protons are typically insufficient to explain the cosmic-ray measurements unless the neutrino output is suppressed.  The first important implication is that, if the source conditions do not include a mechanism for effective pion/muon cooling, the dominant class of accelerator should allow for proton escape without significant neutrino production.  This might also imply the benefit of allowing for efficient escape of nuclei, which tend to be more fragile due to the lower thresholds for photodisintegration.

For instance, increasing the Model {\color{c4} 4} proton flux by a factor of 5 brings good agreement near $10^{18}\,$eV.  At lower energies, the UHECR spectrum is presumably due to Galactic nuclei (see, e.g., \cite{Gaisser:2013bla}).  At $\gtrsim\,$$10^{19}\,$eV, the origin of the received flux is unclear, with a heavy-nuclear composition inferred by Auger \cite{Abraham:2010yv}, in contrast to HiRes \cite{Abbasi:2009nf}, and the possible influence of Cen~A as a local source (see \cite{Yuksel:2012ee}).  If we take the Auger results at face value and assume that nuclei make up the difference, similar to the model of \cite{Aloisio:2009sj}, a transition in composition at $\sim\,$$10^{18.5}\,$eV results, although we have not addressed this in detail.

Dividing our inferred neutrino emissivity for Models {\color{c3} 3} and {\color{c4} 4} into sources with space density $n$ yields an average of $\sim\! 10^{37}\,$erg$\,$s$^{-1}$(Mpc$^{-3}/n$).  This simple energetics requirement can be met by galaxies or a variety of AGN classes, although more exotic scenarios may be considered \cite{Barger:2013pla,Borriello:2013ala,Stecker:2013jfa,Diaz:2013wia,Anchordoqui:2014hua,Stecker:2014xja,Chen:2013dza,Feldstein:2013kka,Esmaili:2013gha}.  The neutrino flux is accompanied by roughly two to four times this energy in gamma rays, which will quickly cascade in or near the source.  Model~{\color{c2} 2} also implies synchrotron gamma rays from charged pions and muons.  Since neutral pions are not cooled, their spectrum would thus extend to a higher energy.  These should be discussed in terms of concrete source models.

Our use of constant cosmic evolution is similar to that of intermediate-luminosity quasars (e.g., \cite{Assef:2010ew}).  Stronger source evolution, such as from bright quasars \cite{Assef:2010ew}, gamma-ray bursts \cite{Kistler:2009mv}, the cosmic star formation history \cite{Hopkins:2006bw}, or black hole production \cite{Yuksel:2012zy}, would result in relatively more neutrinos.  Using the SFR fit from \cite{Kistler:2013jza} would decrease the required per-source neutrino emissivity by a factor of $\sim\!4$, with the spectral peaks shifting to slightly lower energies.  While cosmic rays at the lower end of the energy range would roughly have a corresponding upward shift, at higher energies loss rates are more severe and the spectrum is relatively steepened due to a lack of compensation from more distant sources.  This works in the same direction of our main conclusion relating the sources of cosmic-ray protons and neutrinos.

If a fraction of the IceCube events turn out to be from the Milky Way, the inferred extragalactic neutrino flux would decrease.  This would only strengthen our conclusion that considering photohadronic interactions in isolation is not sufficient to account for freeing cosmic-ray protons from their birthplaces and that neutrino production should be relatively inefficient wherever the cosmic rays arise.  It is also possible that neutrinos are being produced where cosmic rays cannot escape at all (e.g., \cite{Mannheim:1998wp,Stecker:1991vm}).  IceCube may thus be peering into the workings of a class of accelerator distinct from those yielding protons, deepening the mystery of the origin of cosmic rays, while informing us of an aspect of the extreme universe that would not otherwise be evident.

%%---------------------------------------------------------------------%
%\textit{Acknowledgments}--- %
\vspace{-0.35in}
\acknowledgments
\vspace{-0.2in}
We thank John Learned, Sandip Pakvasa, Doug Cowen, Ty DeYoung, and especially Carsten Rott for useful discussions, and the authors of CRPropa for making their code available.
MDK acknowledges support provided by NASA through the Einstein Fellowship Program, grant PF0-110074, by Department of Energy contract DE-AC02-76SF00515, and the KIPAC Kavli Fellowship made possible by The Kavli Foundation; 
TS by DOE Grant DE-FG02-91ER40626;
 and HY by the LANL LDRD program, during a visit to Berkeley by US DOE contract DE-SC00046548, and by The Scientific and Technological Research Council of Turkey (TUBITAK), cofunded by Marie Curie Actions under FP7.

%---------------------------------------------------------------------%
%\textbf{References}


\vspace*{-0.7cm}
\begin{thebibliography}{99}
\vspace*{-1.2cm}

%\cite{Gaisser:1994yf}
\bibitem{Gaisser:1994yf} 
  T.~K.~Gaisser, F.~Halzen and T.~Stanev,
  %``Particle astrophysics with high-energy neutrinos,''
  Phys.\ Rept.\  {\bf 258}, 173 (1995).
  %[Erratum-ibid.\  {\bf 271}, 355 (1996)]
  %[hep-ph/9410384].
  %%CITATION = HEP-PH/9410384;%%

%\cite{Learned:2000sw}
\bibitem{Learned:2000sw} 
  J.~G.~Learned and K.~Mannheim,
  %``High-energy neutrino astrophysics,''
  Ann.\ Rev.\ Nucl.\ Part.\ Sci.\  {\bf 50}, 679 (2000).
  %%CITATION = ARNUA,50,679;%%

%\cite{Halzen:2002pg}
\bibitem{Halzen:2002pg} 
  F.~Halzen and D.~Hooper,
  %``High-energy neutrino astronomy: The Cosmic ray connection,''
  Rept.\ Prog.\ Phys.\  {\bf 65}, 1025 (2002).
  %[astro-ph/0204527].
  %%CITATION = ASTRO-PH/0204527;%%

%\cite{Becker:2007sv}
\bibitem{Becker:2007sv} 
  J.~K.~Becker,
  %``High-energy neutrinos in the context of multimessenger physics,''
  Phys.\ Rept.\  {\bf 458}, 173 (2008).
  %[arXiv:0710.1557 [astro-ph]].
  %%CITATION = ARXIV:0710.1557;%%

%\cite{Roberts:1992re}
\bibitem{Roberts:1992re} 
  A.~Roberts,
  %``The Birth of high-energy neutrino astronomy: A Personal history of the DUMAND project,''
  Rev.\ Mod.\ Phys.\  {\bf 64}, 259 (1992).
  %%CITATION = RMPHA,64,259;%%

%\cite{Halzen:1988wr}
\bibitem{Halzen:1988wr} 
  F.~Halzen and J.~G.~Learned,
  %``High-energy Neutrino Detection In Deep Polar Ice,''
  MAD/PH/428 (1988).
  %Univ.\ of Wisconsin publ.\ MAD/PH/428 (1988).
  %Proceedings of the 5th International Symposium on Very High Energy Cosmic Ray Interactions, Lodz, Poland (1988)
  %%CITATION = MAD/PH/428;%%

%\cite{Barwick:1991ur}
\bibitem{Barwick:1991ur} 
  S.~Barwick, F.~Halzen, D.~Lowder, T.~Miller, R.~Morse, P.~B.~Price and A.~Westphal,
  %``Neutrino astronomy on the 1-KM**2 scale,''
  J.\ Phys.\ G {\bf 18}, 225 (1992).
  %%CITATION = JPHGB,G18,225;%%

%\cite{Ahrens:2002dv}
\bibitem{Ahrens:2002dv} 
  J.~Ahrens {\it et al.}  [IceCube Collaboration],
  %``Icecube - the next generation neutrino telescope at the south pole,''
  Nucl.\ Phys.\ Proc.\ Suppl.\  {\bf 118}, 388 (2003).
  %[astro-ph/0209556].
  %%CITATION = ASTRO-PH/0209556;%%

%\cite{Aartsen:2013bka}
\bibitem{Aartsen:2013bka} 
  M.~G.~Aartsen {\it et al.}  [IceCube Collaboration],
  %``First observation of PeV-energy neutrinos with IceCube,''
  Phys.\ Rev.\ Lett.\  {\bf 111}, 021103 (2013).
  %arXiv:1304.5356.
  %%CITATION = ARXIV:1304.5356;%%

%\cite{Whitehorn}
\bibitem{Whitehorn}
  M.~G.~Aartsen {\it et al.}  [IceCube Collaboration],
  %Evidence for High-Energy Extraterrestrial Neutrinos at the IceCube Detector
  Science {\bf 342}, 1242856 (2013).
  %%CITATION = SCIEA,342,1242856;%%

%\cite{Klein:2013}
\bibitem{Klein:2013} 
  M.~G.~Aartsen {\it et al.}  [IceCube Collaboration],
   %``Observation of High-Energy Astrophysical Neutrinos in Three Years of IceCube Data,''
   Phys.\ Rev.\ Lett.\  {\bf 113}, 101101 (2014).
  %arXiv:1405.5303.
  %%CITATION = ARXIV:1405.5303;%%


%\cite{Gaisser:2002jj}
\bibitem{Gaisser:2002jj} 
  T.~K.~Gaisser and M.~Honda,
  %``Flux of atmospheric neutrinos,''
  Ann.\ Rev.\ Nucl.\ Part.\ Sci.\  {\bf 52}, 153 (2002).
  %[hep-ph/0203272].
  %%CITATION = HEP-PH/0203272;%%

%\cite{Enberg:2008te}
\bibitem{Enberg:2008te} 
  R.~Enberg, M.~H.~Reno and I.~Sarcevic,
  %``Prompt neutrino fluxes from atmospheric charm,''
  Phys.\ Rev.\ D {\bf 78}, 043005 (2008).
  %[arXiv:0806.0418 [hep-ph]].
  %%CITATION = ARXIV:0806.0418;%%


%\cite{Greisen:1966jv}
\bibitem{Greisen:1966jv} 
  K.~Greisen,
  %``End to the cosmic ray spectrum?,''
  Phys.\ Rev.\ Lett.\  {\bf 16}, 748 (1966).
  %%CITATION = PRLTA,16,748;%%

%\cite{Zatsepin:1966jv}
\bibitem{Zatsepin:1966jv} 
  G.~T.~Zatsepin and V.~A.~Kuzmin,
  %``Upper limit of the spectrum of cosmic rays,''
  JETP Lett.\  {\bf 4}, 78 (1966).
  %[Pisma Zh.\ Eksp.\ Teor.\ Fiz.\  {\bf 4}, 114 (1966)].
  %%CITATION = JTPLA,4,78;%%

%\cite{Abbasi:2007sv}
\bibitem{Abbasi:2007sv}
  R.~U.~Abbasi {\it et al.},
  %  [HiRes Collaboration],
  %``First observation of the Greisen-Zatsepin-Kuzmin suppression,''
  Phys.\ Rev.\ Lett.\  {\bf 100}, 101101 (2008).
  %[arXiv:astro-ph/0703099].
  %%CITATION = PRLTA,100,101101;%%

%\cite{Abraham:2008ru}
\bibitem{Abraham:2008ru}
  J.~Abraham {\it et al.},
  %  [Pierre Auger Collaboration],
  %``Observation of the suppression of the flux of cosmic rays above $4\times 10^{19}$eV,''
  Phys.\ Rev.\ Lett.\  {\bf 101}, 061101 (2008).
  %[arXiv:0806.4302 [astro-ph]].
  %%CITATION = PRLTA,101,061101;%%

%\cite{Abreu:2011pj}
\bibitem{Abreu:2011pj} 
  P.~Abreu {\it et al.},
  %  [Pierre Auger Collaboration],
  %``The Pierre Auger Observatory I: The Cosmic Ray Energy Spectrum and Related Measurements,''
  arXiv:1307.5059.
  %%CITATION = ARXIV:1307.5059;%%

%\cite{AbuZayyad:2012ru}
\bibitem{AbuZayyad:2012ru} 
  T.~Abu-Zayyad {\it et al.},
  %, R.~Aida, M.~Allen, R.~Anderson, R.~Azuma, E.~Barcikowski, J.~W.~Belz and D.~R.~Bergman {\it et al.},
  %``The Cosmic Ray Energy Spectrum Observed with the Surface Detector of the Telescope Array Experiment,''
  Astrophys.\ J.\  {\bf 768}, L1 (2013);
  %arXiv:1205.5067.
  %%CITATION = ARXIV:1205.5067;%%
%\cite{Bergman}
%\bibitem{Bergman}
  D.~Bergman  [TA Collaboration],
  %``Precision measurement of the proton flux with AMS,''
  Proc.\ 33rd Intl.\ Cosmic Ray Conf., Rio de Janeiro, {\bf 1}, 0221 (2013).
  %

%\cite{Beresinsky:1969qj}
\bibitem{Beresinsky:1969qj} 
  V.~S.~Berezinsky and G.~T.~Zatsepin,
  %``Cosmic rays at ultrahigh-energies (neutrino?),''
  Phys.\ Lett.\ B {\bf 28}, 423 (1969).
  %%CITATION = PHLTA,B28,423;%%

%\cite{Stecker:1978ah}
\bibitem{Stecker:1978ah} 
  F.~W.~Stecker,
  %``Diffuse Fluxes of Cosmic High-Energy Neutrinos,''
  Astrophys.\ J.\  {\bf 228}, 919 (1979).
  %%CITATION = ASJOA,228,919;%%

%\cite{Hill:1983mk}
\bibitem{Hill:1983mk} 
  C.~T.~Hill and D.~N.~Schramm,
  %``The Ultrahigh-Energy Cosmic Ray Spectrum,''
  Phys.\ Rev.\ D {\bf 31}, 564 (1985).
  %%CITATION = PHRVA,D31,564;%%

%\cite{Yoshida:1993pt}
\bibitem{Yoshida:1993pt} 
  S.~Yoshida and M.~Teshima,
  %``Energy spectrum of ultrahigh-energy cosmic rays with extragalactic origin,''
  Prog.\ Theor.\ Phys.\  {\bf 89}, 833 (1993).
  %%CITATION = PTPKA,89,833;%%

%\cite{Waxman:1998yy}
\bibitem{Waxman:1998yy} 
  E.~Waxman and J.~N.~Bahcall,
  %``High-energy neutrinos from astrophysical sources: An Upper bound,''
  Phys.\ Rev.\ D {\bf 59}, 023002 (1999).
  %[hep-ph/9807282].
  %%CITATION = HEP-PH/9807282;%%

%\cite{Engel:2001hd}
\bibitem{Engel:2001hd}
  R.~Engel, D.~Seckel and T.~Stanev,
  %``Neutrinos from propagation of ultra-high energy protons,''
  Phys.\ Rev.\  D {\bf 64}, 093010 (2001).
  %[arXiv:astro-ph/0101216].
  %%CITATION = PHRVA,D64,093010;%%


%\cite{Stanev:2004kz}
\bibitem{Stanev:2004kz}
  T.~Stanev,
  %``Neutrino production in UHECR proton interactions in the infrared background,''
  Phys.\ Lett.\ B {\bf 595}, 50 (2004).
  %[astro-ph/0404535].
  %%CITATION = ASTRO-PH/0404535;%%

%\cite{Bugaev:2004xt}
\bibitem{Bugaev:2004xt} 
  E.~V.~Bugaev, A.~Misaki and K.~Mitsui,
  %``Neutrinos from extragalactic cosmic ray interactions in the far infrared background,''
  Astropart.\ Phys.\  {\bf 24}, 345 (2005).
  %[astro-ph/0405109].
  %%CITATION = ASTRO-PH/0405109;%%

%\cite{DeMarco:2005kt}
\bibitem{DeMarco:2005kt}
  T.~Stanev, D.~De Marco, M.~A.~Malkan and F.~W.~Stecker,
  %``Cosmogenic neutrinos from cosmic ray interactions with extragalactic infrared photons,''
  Phys.\ Rev.\ D {\bf 73}, 043003 (2006).
  %[astro-ph/0512479].
  %%CITATION = ASTRO-PH/0512479;%%

%\cite{Ackermann:2012}
\bibitem{Ackermann:2012} 
  M.~Ackermann {\it et al.},
  %``The Imprint of the Extragalactic Background Light in the Gamma-Ray Spectra of Blazars,''
  Science {\bf 338}, 1190 (2012).
  %%CITATION = SCIEA,338,1190;%%

%\cite{Roulet}
\bibitem{Roulet}
  E.~Roulet, G.~Sigl, A.~van Vliet and S.~Mollerach,
  %``PeV neutrinos from the propagation of ultra-high energy cosmic rays,''
  JCAP {\bf 1301}, 028 (2013).
  %arXiv:1209.4033.
  %%CITATION = ARXIV:1209.4033;%%


%\cite{Abbasi:2012zw}
\bibitem{Abbasi:2012zw} 
  R.~Abbasi {\it et al.}  [IceCube Collaboration],
  %``An absence of neutrinos associated with cosmic-ray acceleration in $\gamma$-ray bursts,''
  Nature {\bf 484}, 351 (2012).
  %[arXiv:1204.4219 [astro-ph.HE]].
  %%CITATION = ARXIV:1204.4219;%%

%\cite{Hummer:2011ms}
\bibitem{Hummer:2011ms} 
  S.~Hummer, P.~Baerwald and W.~Winter,
  %``Neutrino Emission from Gamma-Ray Burst Fireballs, Revised,''
  Phys.\ Rev.\ Lett.\  {\bf 108}, 231101 (2012).
  %[arXiv:1112.1076 [astro-ph.HE]].
  %%CITATION = ARXIV:1112.1076;%%

%\cite{Cholis:2012kq}
\bibitem{Cholis:2012kq} 
  I.~Cholis and D.~Hooper,
  %``On The Origin of IceCube's PeV Neutrinos,''
  JCAP {\bf 06}, 030 (2013).
  %arXiv:1211.1974.
  %%CITATION = ARXIV:1211.1974;%%

%\cite{Liu:2012pf}
\bibitem{Liu:2012pf} 
  R.~-Y.~Liu and X.~-Y.~Wang,
  %``Diffuse PeV neutrinos from gamma-ray bursts,''
  Astrophys.\ J.\  {\bf 766}, 73 (2013).
  %arXiv:1212.1260.
  %%CITATION = ARXIV:1212.1260;%%

%\cite{Baerwald:2014zga}
\bibitem{Baerwald:2014zga} 
  P.~Baerwald, M.~Bustamante and W.~Winter,
  %``Are gamma-ray bursts the sources of ultra-high energy cosmic rays?,''
  Astropart.\ Phys.\  {\bf 62}, 66 (2014).
  %arXiv:1401.1820.
  %%CITATION = ARXIV:1401.1820;%%


%\cite{Hillas:1985is}
\bibitem{Hillas:1985is} 
  A.~M.~Hillas,
  %``The Origin of Ultrahigh-Energy Cosmic Rays,''
  Ann.\ Rev.\ Astron.\ Astrophys.\  {\bf 22}, 425 (1984).
  %%CITATION = ARAAA,22,425;%%


%\cite{Abbasi:2004nz}
\bibitem{Abbasi:2004nz} 
  R.~U.~Abbasi {\it et al.},
  %  [HiRes Collaboration],
  %``A Study of the composition of ultrahigh energy cosmic rays using the High Resolution Fly's Eye,''
  Astrophys.\ J.\  {\bf 622}, 910 (2005).
  %[astro-ph/0407622].
  %%CITATION = ASTRO-PH/0407622;%%

%\cite{Abbasi:2009nf}
\bibitem{Abbasi:2009nf} 
  R.~U.~Abbasi {\it et al.},
  %  [HiRes Collaboration],
  %``Indications of Proton-Dominated Cosmic Ray Composition above 1.6 EeV,''
  Phys.\ Rev.\ Lett.\  {\bf 104}, 161101 (2010).
  %[arXiv:0910.4184 [astro-ph.HE]].
  %%CITATION = ARXIV:0910.4184;%%

%\cite{Abraham:2010yv}
\bibitem{Abraham:2010yv} 
  J.~Abraham {\it et al.},
  %  [Pierre Auger Collaboration],
  %``Measurement of the Depth of Maximum of Extensive Air Showers above 10^18 eV,''
  Phys.\ Rev.\ Lett.\  {\bf 104}, 091101 (2010).
  %[arXiv:1002.0699 [astro-ph.HE]].
  %%CITATION = ARXIV:1002.0699;%%

%\cite{Apel:2013dga}
\bibitem{Apel:2013dga} 
  W.~D.~Apel, {\it et al.},
  %W.~D.~Apel, J.~C.~Arteaga-Vel‡zquez, K.~Bekk, {\it et al.},
  %``KASCADE-Grande measurements of energy spectra for elemental groups of cosmic rays,''
  Astropart.\ Phys.\  {\bf 47}, 54 (2013).
  %[arXiv:1306.6283 [astro-ph.HE]].
  %%CITATION = ARXIV:1306.6283;%%

%\cite{Barcikowski:2013nfa}
\bibitem{Barcikowski:2013nfa} 
  E.~Barcikowski {\it et al.} % [Auger M. Unger for the Pierre and Yakutsk Collaborations],
  %``Mass Composition Working Group Report at UHECR-2012,''
  EPJ Web Conf.\  {\bf 53}, 01006 (2013).
  %[arXiv:1306.4430 [astro-ph.HE]].
  %%CITATION = ARXIV:1306.4430;%%


%\cite{Pakvasa:2012db}
\bibitem{Pakvasa:2012db} 
  S.~Pakvasa, A.~Joshipura and S.~Mohanty,
  %``Explanation for the low flux of high energy astrophysical muon-neutrinos,''
  Phys.\ Rev.\ Lett.\  {\bf 110}, 171802 (2013).
  %[arXiv:1209.5630 [hep-ph]].
  %%CITATION = ARXIV:1209.5630;%%

%\cite{Kalashev:2013vba}
\bibitem{Kalashev:2013vba} 
  O.~E.~Kalashev, A.~Kusenko and W.~Essey,
  %``PeV neutrinos from intergalactic interactions of cosmic rays emitted by active galactic nuclei,''
  Phys.\ Rev.\ Lett.\  {\bf 111}, 041103 (2013).
  %[arXiv:1303.0300 [astro-ph.HE]].
  %%CITATION = ARXIV:1303.0300;%%

%\cite{Arsene:2013nca}
\bibitem{Arsene:2013nca} 
  N.~Arsene, X.~Calmet, L.~I.~Caramete and O.~Micu,
  %``Back-to-Back Black Holes decay Signature at Neutrino Observatories,''
  Astropart.\ Phys.\  {\bf 54}, 132 (2014).
  %[arXiv:1303.4603 [hep-ph]].
  %%CITATION = ARXIV:1303.4603;%%

%\cite{Fox:2013oza}
\bibitem{Fox:2013oza} 
  D.~B.~Fox, K.~Kashiyama and P.~Meszaros,
  %``Sub-PeV Neutrinos from TeV Unidentified Sources in the Galaxy,''
  Astrophys.\ J.\  {\bf 774}, 74 (2013).
  %[arXiv:1305.6606 [astro-ph.HE]].
  %%CITATION = ARXIV:1305.6606;%%

%\cite{Vissani:2013iga}
\bibitem{Vissani:2013iga} 
  F.~Vissani, G.~Pagliaroli and F.~L.~Villante,
  %``The fraction of muon tracks in cosmic neutrinos,''
  JCAP {\bf 1309}, 017 (2013).
  %[arXiv:1306.0211 [astro-ph.HE]].
  %%CITATION = ARXIV:1306.0211;%%

%\cite{Laha:2013lka}
\bibitem{Laha:2013lka} 
  R.~Laha, J.~F.~Beacom, B.~Dasgupta, S.~Horiuchi and K.~Murase,
  %``Demystifying the PeV Cascades in IceCube: Less (Energy) is More (Events),''
  Phys.\ Rev.\ D {\bf 88}, 043009 (2013).
  %[arXiv:1306.2309 [astro-ph.HE]].
  %%CITATION = ARXIV:1306.2309;%%

%\cite{Murase:2013rfa}
\bibitem{Murase:2013rfa} 
  K.~Murase, M.~Ahlers and B.~C.~Lacki,
  %``Testing the Hadronuclear Origin of PeV Neutrinos Observed with IceCube,''
  Phys.\ Rev.\ D {\bf 88}, 121301 (2013).
  %arXiv:1306.3417.
  %%CITATION = ARXIV:1306.3417;%%

%\cite{Anchordoqui:2013qsi}
\bibitem{Anchordoqui:2013qsi} 
  L.~A.~Anchordoqui, H.~Goldberg, M.~H.~Lynch, A.~V.~Olinto, T.~C.~Paul and T.~J.~Weiler,
  %``Pinning down the cosmic ray source mechanism with new IceCube data,''
  Phys.\ Rev.\ D {\bf 89}, 083003 (2014).
  %[arXiv:1306.5021 [astro-ph.HE]].
  %%CITATION = ARXIV:1306.5021;%%

%\cite{Winter:2013cla}
\bibitem{Winter:2013cla} 
  W.~Winter,
  %``Photohadronic Origin of the TeV-PeV Neutrinos Observed in IceCube,''
  Phys.\ Rev.\ D {\bf 88}, 083007 (2013).
  %[arXiv:1307.2793 [astro-ph.HE]].
  %%CITATION = ARXIV:1307.2793;%%

%\cite{Ahlers:2013xia}
\bibitem{Ahlers:2013xia} 
  M.~Ahlers and K.~Murase,
  %``Probing the Galactic Origin of the IceCube Excess with Gamma-Rays,''
  Phys.\ Rev.\ D {\bf 90}, 023010 (2014).
  %arXiv:1309.4077.
  %%CITATION = ARXIV:1309.4077;%%

%\cite{Halzen:2013dva}
\bibitem{Halzen:2013dva} 
  F.~Halzen,
  %``The highest energy neutrinos: first evidence for cosmic origin,''
  Astron.\ Nachr.\  {\bf 335}, 507 (2014).
  %arXiv:1311.6350.
  %%CITATION = ARXIV:1311.6350;%%

%\cite{Anchordoqui:2013dnh}
\bibitem{Anchordoqui:2013dnh} 
  L.~A.~Anchordoqui, V.~Barger, I.~Cholis, H.~Goldberg, D.~Hooper, A.~Kusenko, J.~G.~Learned, D.~Marfatia, S.~Pakvasa, T.~C.~Paul and T.~J.~Weiler,
  %``Cosmic Neutrino Pevatrons: A Brand New Pathway to Astronomy, Astrophysics, and Particle Physics,''
  Journal of High Energy Astrophysics {\bf 1-2}, 1 (2014).
  %[arXiv:1312.6587 [astro-ph.HE]].
  %%CITATION = ARXIV:1312.6587;%%

%\cite{Fang:2014uja}
\bibitem{Fang:2014uja} 
  K.~Fang, T.~Fujii, T.~Linden and A.~V.~Olinto,
  %``Is the Ultra-High Energy Cosmic-Ray Excess Observed by the Telescope Array Correlated with IceCube Neutrinos?,''
  arXiv:1404.6237.
  %%CITATION = ARXIV:1404.6237;%%

%\cite{Kachelriess:2014oma}
\bibitem{Kachelriess:2014oma} 
  M.~Kachelriess and S.~Ostapchenko,
  %``Neutrino yield from Galactic cosmic rays,''
  Phys.\ Rev.\ D {\bf 90}, 083002 (2014).
  %arXiv:1405.3797.
  %%CITATION = ARXIV:1405.3797;%%

%\cite{Learned:2014vya}
\bibitem{Learned:2014vya} 
  J.~G.~Learned and T.~J.~Weiler,
  %``A Relational Argument for a $\sim$PeV Neutrino Energy Cutoff,''
  arXiv:1407.0739.
  %%CITATION = ARXIV:1407.0739;%%

%\cite{Winter:2014pya}
\bibitem{Winter:2014pya} 
  W.~Winter,
  %``Describing the Observed Cosmic Neutrinos by Interactions of Nuclei with Matter,''
  arXiv:1407.7536.
  %%CITATION = ARXIV:1407.7536;%%


%\cite{Mannheim:1998wp}
\bibitem{Mannheim:1998wp} 
  K.~Mannheim, R.~J.~Protheroe and J.~P.~Rachen,
  %``On the cosmic ray bound for models of extragalactic neutrino production,''
  Phys.\ Rev.\ D {\bf 63}, 023003 (2001).
  %[astro-ph/9812398].
  %%CITATION = ASTRO-PH/9812398;%%



%\cite{Klein:2012ug}
\bibitem{Klein:2012ug} 
  S.~R.~Klein, R.~Mikkelsen and J.~K.~B.~Tjus,
  %``Muon Acceleration in Cosmic-ray Sources,''
  Astrophys.\ J.\  {\bf 779}, 106 (2013).
  %[arXiv:1208.2056 [astro-ph.HE]].
  %%CITATION = ARXIV:1208.2056;%%

%\cite{Rachen:1998fd}
\bibitem{Rachen:1998fd} 
  J.~P.~Rachen and P.~Meszaros,
  %``Photohadronic neutrinos from transients in astrophysical sources,''
  Phys.\ Rev.\ D {\bf 58}, 123005 (1998).
  %[astro-ph/9802280].
  %%CITATION = ASTRO-PH/9802280;%%

%\cite{Winter:2012xq}
\bibitem{Winter:2012xq} 
  W.~Winter,
  %``Neutrinos from Cosmic Accelerators Including Magnetic Field and Flavor Effects,''
  Adv.\ High Energy Phys.\  {\bf 2012}, 586413 (2012).
  %[arXiv:1201.5462 [astro-ph.HE]].
  %%CITATION = ARXIV:1201.5462;%%


%\cite{Beringer:1900zz}
\bibitem{Beringer:1900zz} 
  J.~Beringer {\it et al.},
  %  [Particle Data Group Collaboration],
  %``Review of Particle Physics (RPP),''
  Phys.\ Rev.\ D {\bf 86}, 010001 (2012).
  %%CITATION = PHRVA,D86,010001;%%


%\cite{Gandhi:1998ri}
\bibitem{Gandhi:1998ri}
  R.~Gandhi, C.~Quigg, M.~H.~Reno and I.~Sarcevic,
  %``Neutrino interactions at ultrahigh energies,''
  Phys.\ Rev.\  D {\bf 58}, 093009 (1998).
  %%CITATION = PHRVA,D58,093009;%%

%\cite{Gandhi:1995tf}
\bibitem{Gandhi:1995tf}
  R.~Gandhi, C.~Quigg, M.~H.~Reno and I.~Sarcevic,
  %``Ultrahigh-energy neutrino interactions,''
  Astropart.\ Phys.\  {\bf 5}, 81 (1996).
  %%CITATION = APHYE,5,81;%%

%\cite{Kowalski:2004qc}
\bibitem{Kowalski:2004qc}
  M.~P.~Kowalski,
  Ph.D.\ thesis (Humboldt, 2004).
  %``Search for neutrino induced cascades with the AMANDA-II detector,''
  %\href{http://www.slac.stanford.edu/spires/find/hep/www?irn=5826756}{SPIRES entry}
  %%CITATION = INSPIRE-646563;%%

%\cite{Aartsen:2013vja}
\bibitem{Aartsen:2013vja} 
  M.~G.~Aartsen {\it et al.},
  %  [IceCube Collaboration],
  %``Energy Reconstruction Methods in the IceCube Neutrino Telescope,''
  JINST {\bf 9}, P03009 (2014).
  %[arXiv:1311.4767 [physics.ins-det]].
  %%CITATION = ARXIV:1311.4767;%%


%\cite{Learned:1994wg}
\bibitem{Learned:1994wg} 
  J.~G.~Learned and S.~Pakvasa,
  %``Detecting tau-neutrino oscillations at PeV energies,''
  Astropart.\ Phys.\  {\bf 3}, 267 (1995).
  %[hep-ph/9405296].
  %%CITATION = HEP-PH/9405296,;%%

%\cite{Gaisser:1990vg}
\bibitem{Gaisser:1990vg}
  T.~K. Gaisser, \textit{Cosmic Rays and Particle Physics},
  (Cambridge Univ. Press, Cambridge, 1990).

%\cite{Kistler:2006hp}
\bibitem{Kistler:2006hp}
  M.~D.~Kistler and J.~F.~Beacom,
  %``Guaranteed and prospective galactic TeV neutrino sources,''
  Phys.\ Rev.\ D {\bf 74}, 063007 (2006).
  %[arXiv:astro-ph/0607082].
  %%CITATION = ASTRO-PH 0607082;%%

%\cite{Beacom:2007yu}
\bibitem{Beacom:2007yu} 
  J.~F.~Beacom and M.~D.~Kistler,
  %``Dissecting the Cygnus Region with TeV Gamma Rays and Neutrinos,''
  Phys.\ Rev.\ D {\bf 75}, 083001 (2007).
  %[astro-ph/0701751].
  %%CITATION = ASTRO-PH/0701751;%%

%\cite{Dziewonski:1981xy}
\bibitem{Dziewonski:1981xy}
  A.~M.~Dziewonski and D.~L.~Anderson,
  %``Preliminary reference earth model,''
  Phys.\ Earth Planet.\ Interiors {\bf 25}, 297 (1981).
  %%CITATION = PEPIA,25,297;%%


%\cite{Aartsen:2013eka}
\bibitem{Aartsen:2013eka} 
  M.~G.~Aartsen {\it et al.}  [IceCube Collaboration],
  %``Search for a diffuse flux of astrophysical muon neutrinos with the IceCube 59-string configuration,''
  Phys.\ Rev.\ D {\bf 89}, 062007 (2014).
  %[arXiv:1311.7048 [astro-ph.HE]].
  %%CITATION = ARXIV:1311.7048;%%


%\cite{Beacom:2003nh}
\bibitem{Beacom:2003nh} 
  J.~F.~Beacom, N.~F.~Bell, D.~Hooper, S.~Pakvasa and T.~J.~Weiler,
  %``Measuring flavor ratios of high-energy astrophysical neutrinos,''
  Phys.\ Rev.\ D {\bf 68}, 093005 (2003).
  %[Erratum-ibid.\ D {\bf 72}, 019901 (2005)]
  %[hep-ph/0307025].
  %%CITATION = HEP-PH/0307025;%%

%\cite{Anchordoqui:2004eb}
\bibitem{Anchordoqui:2004eb} 
  L.~A.~Anchordoqui, H.~Goldberg, F.~Halzen and T.~J.~Weiler,
  %``Neutrinos as a diagnostic of high energy astrophysical processes,''
  Phys.\ Lett.\ B {\bf 621}, 18 (2005).
  %[hep-ph/0410003].
  %%CITATION = HEP-PH/0410003;%%

%\cite{Barger:2012mz}
\bibitem{Barger:2012mz} 
  V.~Barger, J.~Learned and S.~Pakvasa,
  %``IceCube Neutrino Initiated Cascade Events: PeV Electron-antineutrinos at Glashow Resonance,''
  Phys.\ Rev.\ D {\bf 87}, 037302 (2013).
  %arXiv:1207.4571.
  %%CITATION = ARXIV:1207.4571;%%

%\cite{Bhattacharya:2012fh}
\bibitem{Bhattacharya:2012fh} 
  A.~Bhattacharya, R.~Gandhi, W.~Rodejohann and A.~Watanabe,
  %``On the interpretation of IceCube cascade events in terms of the Glashow resonance,''
  arXiv:1209.2422.
  %%CITATION = ARXIV:1209.2422;%%

%\cite{Barger:2014iua}
\bibitem{Barger:2014iua} 
  V.~Barger, L.~Fu, J.~G.~Learned, D.~Marfatia, S.~Pakvasa and T.~J.~Weiler,
  %``Glashow resonance as a window into cosmic neutrino sources,''
  arXiv:1407.3255.
  %%CITATION = ARXIV:1407.3255;%%

%\cite{Lipari:1991ut}
\bibitem{Lipari:1991ut} 
  P.~Lipari and T.~Stanev,
  %``Propagation of multi - TeV muons,''
  Phys.\ Rev.\ D {\bf 44}, 3543 (1991).
  %%CITATION = PHRVA,D44,3543;%%



%\cite{Aharonian:2000pv}
\bibitem{Aharonian:2000pv} 
  F.~A.~Aharonian,
  %``TeV gamma-rays from BL Lac objects due to synchrotron radiation of extremely high-energy protons,''
  New Astron.\  {\bf 5}, 377 (2000).
  %[astro-ph/0003159].
  %%CITATION = ASTRO-PH/0003159;%%

%\cite{Mucke:1998mk}
\bibitem{Mucke:1998mk}
  A.~Mucke, J.~P.~Rachen, R.~Engel, R.~J.~Protheroe and T.~Stanev,
  %``On photohadronic processes in astrophysical environments,''
  Publ.\ Astron.\ Soc.\ Austral.\  {\bf 16}, 160 (1999).
  %%CITATION = ASTRO-PH 9808279;%%
  
%\cite{Blumenthal:1970nn}
\bibitem{Blumenthal:1970nn}
  G.~R.~Blumenthal,
   %``Energy loss of high-energy cosmic rays in pair-producing collisions with ambient photons,''
  Phys.\ Rev.\ D {\bf 1}, 1596 (1970).
  %%CITATION = PHRVA,D1,1596;%%

%\cite{Berezinsky:2002nc}
\bibitem{Berezinsky:2002nc} 
  V.~Berezinsky, A.~Z.~Gazizov and S.~I.~Grigorieva,
  %``On astrophysical solution to ultrahigh-energy cosmic rays,''
  Phys.\ Rev.\ D {\bf 74}, 043005 (2006).
  %[hep-ph/0204357].
  %%CITATION = HEP-PH/0204357;%%

%\cite{Yuksel:2006qb}
\bibitem{Yuksel:2006qb}
  H.~Yuksel and M.~D.~Kistler,
  %``Enhanced Cosmological GRB Rates and Implications for Cosmogenic
  %Neutrinos,''
  Phys.\ Rev.\  D {\bf 75}, 083004 (2007).
  %[arXiv:astro-ph/0610481].
  %%CITATION = PHRVA,D75,083004;%%

%\cite{Gehrels:1986mj}
\bibitem{Gehrels:1986mj} 
  N.~Gehrels,
  %``Confidence limits for small numbers of events in astrophysical data,''
  Astrophys.\ J.\  {\bf 303}, 336 (1986).
  %%CITATION = ASJOA,303,336;%%

%\cite{Kampert:2012fi}
\bibitem{Kampert:2012fi} 
  K.-H.~Kampert, J.~Kulbartz, L.~Maccione, N.~Nierstenhoefer, P.~Schiffer, G.~Sigl and A.~R.~van Vliet,
  %``CRPropa 2.0 -- a Public Framework for Propagating High Energy Nuclei, Secondary Gamma Rays and Neutrinos,''
  Astropart.\ Phys.\  {\bf 42}, 41 (2013).
  %arXiv:1206.3132.
  %%CITATION = ARXIV:1206.3132;%%

%\cite{Abdo:2010nz}
\bibitem{Abdo:2010nz} 
  A.~A.~Abdo {\it et al.},
  %  [Fermi-LAT Collaboration],
  %``The Spectrum of the Isotropic Diffuse Gamma-Ray Emission Derived From First-Year Fermi Large Area Telescope Data,''
  Phys.\ Rev.\ Lett.\  {\bf 104}, 101101 (2010).
  %[arXiv:1002.3603 [astro-ph.HE]].
  %%CITATION = ARXIV:1002.3603;%%


%\cite{Berezinsky:2010xa}
\bibitem{Berezinsky:2010xa} 
  V.~Berezinsky, A.~Gazizov, M.~Kachelriess and S.~Ostapchenko,
  %``Restricting UHECRs and cosmogenic neutrinos with Fermi-LAT,''
  Phys.\ Lett.\ B {\bf 695}, 13 (2011).
  %[arXiv:1003.1496 [astro-ph.HE]].
  %%CITATION = ARXIV:1003.1496;%%

%\cite{Ahlers:2010fw}
\bibitem{Ahlers:2010fw} 
  M.~Ahlers, L.~A.~Anchordoqui, M.~C.~Gonzalez-Garcia, F.~Halzen and S.~Sarkar,
  %``GZK Neutrinos after the Fermi-LAT Diffuse Photon Flux Measurement,''
  Astropart.\ Phys.\  {\bf 34}, 106 (2010).
  %[arXiv:1005.2620 [astro-ph.HE]].
  %%CITATION = ARXIV:1005.2620;%%

%\cite{Decerprit:2011qe}
\bibitem{Decerprit:2011qe} 
  G.~Decerprit and D.~Allard,
  %``Constraints on the origin of ultrahigh energy cosmic rays from cosmogenic neutrinos and photons,''
  Astron.\ Astrophys.\  {\bf 535}, A66 (2011).
  %[arXiv:1107.3722 [astro-ph.HE]].
  %%CITATION = ARXIV:1107.3722;%%

%\cite{Gelmini:2011kg}
\bibitem{Gelmini:2011kg} 
  G.~B.~Gelmini, O.~Kalashev and D.~V.~Semikoz,
  %``Gamma-Ray Constraints on Maximum Cosmogenic Neutrino Fluxes and UHECR Source Evolution Models,''
  JCAP {\bf 1201}, 044 (2012).
  %[arXiv:1107.1672 [astro-ph.CO]].
  %%CITATION = ARXIV:1107.1672;%%

%\cite{Murase:2012df}
\bibitem{Murase:2012df} 
  K.~Murase, J.~F.~Beacom and H.~Takami,
  %``Gamma-Ray and Neutrino Backgrounds as Probes of the High-Energy Universe: Hints of Cascades, General Constraints, and Implications for TeV Searches,''
  JCAP {\bf 1208}, 030 (2012).
  %arXiv:1205.5755.
  %%CITATION = ARXIV:1205.5755;%%



%\cite{Ahlers:2009rf}
\bibitem{Ahlers:2009rf} 
  M.~Ahlers, L.~A.~Anchordoqui and S.~Sarkar,
  %``Neutrino diagnostics of ultra-high energy cosmic ray protons,''
  Phys.\ Rev.\ D {\bf 79}, 083009 (2009).
  %[arXiv:0902.3993 [astro-ph.HE]].
  %%CITATION = ARXIV:0902.3993;%%

%\cite{O'Sullivan:2009ua}
\bibitem{O'Sullivan:2009ua}
  S.~P.~O'Sullivan and D.~C.~Gabuzda,
  %``Magnetic field strength and spectral distribution of six parsec-scale active galactic nuclei jets,''
  Mon.\ Not.\ Roy.\ Astron.\ Soc.\ {\bf 400}, 26 (2009).
  %arXiv:0907.5211 [astro-ph.CO].
  %%CITATION = ARXIV:0907.5211;%%

%\cite{Stecker:1991vm}
\bibitem{Stecker:1991vm} 
  F.~W.~Stecker, C.~Done, M.~H.~Salamon and P.~Sommers,
  %``High-energy neutrinos from active galactic nuclei,''
  Phys.\ Rev.\ Lett.\  {\bf 66}, 2697 (1991).
  %[Erratum-ibid.\  {\bf 69}, 2738 (1992)].
  %%CITATION = PRLTA,66,2697;%%

%\cite{Stecker:2005hn}
\bibitem{Stecker:2005hn} 
  F.~W.~Stecker,
  %``A note on high energy neutrinos from agn cores,''
  Phys.\ Rev.\ D {\bf 72}, 107301 (2005).
  %[astro-ph/0510537].
  %%CITATION = ASTRO-PH/0510537;%%

%\cite{Stecker:2013fxa}
\bibitem{Stecker:2013fxa} 
  F.~W.~Stecker,
  %``PeV neutrinos observed by IceCube from cores of active galactic nuclei,''
  Phys.\ Rev.\ D {\bf 88}, 047301 (2013).
  %[arXiv:1305.7404 [astro-ph.HE]].
  %%CITATION = ARXIV:1305.7404;%%

%\cite{Baerwald:2013pu}
\bibitem{Baerwald:2013pu} 
  P.~Baerwald, M.~Bustamante and W.~Winter,
  %``UHECR escape mechanisms for protons and neutrons from GRBs, and the cosmic ray-neutrino connection,''
  Astrophys.\ J.\  {\bf 768}, 186 (2013).
  %[arXiv:1301.6163 [astro-ph.HE]].
  %%CITATION = ARXIV:1301.6163;%%

%\cite{Petropoulou:2014lja}
\bibitem{Petropoulou:2014lja} 
  M.~Petropoulou, D.~Giannios and S.~Dimitrakoudis,
  %``What IceCube neutrinos teach us about the GRB location,''
  Mon.\ Not.\ Roy.\ Astron.\ Soc.\  {\bf 445}, 570 (2014).
  %arXiv:1405.2091.
  %%CITATION = ARXIV:1405.2091;%%



%\cite{Muecke:2002bi}
\bibitem{Muecke:2002bi} 
  A.~Muecke, R.~J.~Protheroe, R.~Engel, J.~P.~Rachen and T.~Stanev,
  %``BL Lac Objects in the synchrotron proton blazar model,''
  Astropart.\ Phys.\  {\bf 18}, 593 (2003).
  %[astro-ph/0206164].
  %%CITATION = ASTRO-PH/0206164;%%

%\cite{Atoyan:2002gu}
\bibitem{Atoyan:2002gu} 
  A.~M.~Atoyan and C.~D.~Dermer,
  %``Neutral beams from blazar jets,''
  Astrophys.\ J.\  {\bf 586}, 79 (2003).
  %[astro-ph/0209231].
  %%CITATION = ASTRO-PH/0209231;%%

%\cite{Dimitrakoudis:2012uf}
\bibitem{Dimitrakoudis:2012uf} 
  S.~Dimitrakoudis, A.~Mastichiadis, R.~J.~Protheroe and A.~Reimer,
  %``The time-dependent one-zone hadronic model - First principles,''
  Astron.\ Astrophys.\ {\bf 546}, A120 (2012).
  %arXiv:1209.0413.
  %%CITATION = ARXIV:1209.0413;%%

%\cite{Dimitrakoudis:2013tpa}
\bibitem{Dimitrakoudis:2013tpa} 
  S.~Dimitrakoudis, M.~Petropoulou and A.~Mastichiadis,
  %``Self-consistent neutrino and UHE cosmic ray spectra from Mrk 421,''
  Astropart.\ Phys.\  {\bf 54}, 61 (2014).
  %arXiv:1310.7923.
  %%CITATION = ARXIV:1310.7923;%%


%\cite{Gaisser:2013bla}
\bibitem{Gaisser:2013bla} 
  T.~K.~Gaisser, T.~Stanev and S.~Tilav,
  %``Cosmic Ray Energy Spectrum from Measurements of Air Showers,''
  Front.\ Phys.\ China.\  {\bf 8}, 748 (2013).
  %[arXiv:1303.3565 [astro-ph.HE]].
  %%CITATION = ARXIV:1303.3565;%%


%\cite{Yuksel:2012ee}
\bibitem{Yuksel:2012ee} 
  H.~Yuksel, T.~Stanev, M.~D.~Kistler and P.~P.~Kronberg,
  %``The Centaurus A Ultrahigh-Energy Cosmic Ray Excess and the Local Extragalactic Magnetic Field,''
  Astrophys.\ J.\  {\bf 758}, 16 (2012).
  %arXiv:1203.3197.
  %%CITATION = ARXIV:1203.3197;%%


%\cite{Aloisio:2009sj}
\bibitem{Aloisio:2009sj} 
  R.~Aloisio, V.~Berezinsky and A.~Gazizov,
  %``Ultra High Energy Cosmic Rays: The disappointing model,''
  Astropart.\ Phys.\  {\bf 34}, 620 (2011).
  %[arXiv:0907.5194 [astro-ph.HE]].
  %%CITATION = ARXIV:0907.5194;%%


%\cite{Barger:2013pla}
\bibitem{Barger:2013pla} 
  V.~Barger and W.~-Y.~Keung,
  %``Superheavy Particle Origin of IceCube PeV Neutrino Events,''
  Phys.\ Lett.\ B {\bf } (2013).
  %[arXiv:1305.6907 [hep-ph]].
  %%CITATION = ARXIV:1305.6907;%%

%\cite{Borriello:2013ala}
\bibitem{Borriello:2013ala} 
  E.~Borriello, S.~Chakraborty, A.~Mirizzi and P.~D.~Serpico,
  %``Stringent constraint on neutrino Lorentz-invariance violation from the two IceCube PeV neutrinos,''
  Phys.\ Rev.\ D {\bf 87}, 116009 (2013).
  %[arXiv:1303.5843 [astro-ph.HE]].
  %%CITATION = ARXIV:1303.5843;%%

%\cite{Stecker:2013jfa}
\bibitem{Stecker:2013jfa} 
  F.~W.~Stecker,
  %``Constraining Superluminal Electron and Neutrino Velocities using the 2010 Crab Nebula Flare and the IceCube PeV Neutrino Events,''
  Astropart.\ Phys.\  {\bf 56}, 16 (2014).
  %arXiv:1306.6095.
  %%CITATION = ARXIV:1306.6095;%%

%\cite{Diaz:2013wia}
\bibitem{Diaz:2013wia} 
  J.~S.~Diaz, A.~Kostelecky and M.~Mewes,
  %``Testing Relativity with High-Energy Astrophysical Neutrinos,''
  Phys.\ Rev.\ D {\bf 89}, 043005 (2014).
  %arXiv:1308.6344.
  %%CITATION = ARXIV:1308.6344;%%

%\cite{Anchordoqui:2014hua}
\bibitem{Anchordoqui:2014hua} 
  L.~A.~Anchordoqui, V.~Barger, H.~Goldberg, J.~G.~Learned, D.~Marfatia, S.~Pakvasa, T.~C.~Paul and T.~J.~Weiler,
  %``End of the cosmic neutrino energy spectrum,''
  Phys.\ Lett.\ B {\bf 739}, 99 (2014).
  %arXiv:1404.0622.
  %%CITATION = ARXIV:1404.0622;%%

%\cite{Stecker:2014xja}
\bibitem{Stecker:2014xja} 
  F.~W.~Stecker and S.~T.~Scully,
  %``Propagation of Superluminal PeV IceCube Neutrinos: A High Energy Spectral Cutoff or New Constraints on Lorentz Invariance Violation,''
  Phys.\ Rev.\ D {\bf 90}, 043012 (2014).
  %arXiv:1404.7025.
  %%CITATION = ARXIV:1404.7025;%%

%\cite{Chen:2013dza}
\bibitem{Chen:2013dza} 
  C.~-Y.~Chen, P.~S.~B.~Dev and A.~Soni,
  %``Standard Model Explanation of the Ultra-high Energy Neutrino Events at IceCube,''
  Phys.\ Rev.\ D {\bf 89}, 033012 (2014).
  %[arXiv:1309.1764 [hep-ph]].
  %%CITATION = ARXIV:1309.1764;%%

%\cite{Feldstein:2013kka}
\bibitem{Feldstein:2013kka} 
  B.~Feldstein, A.~Kusenko, S.~Matsumoto and T.~T.~Yanagida,
  %``Neutrinos at IceCube from Heavy Decaying Dark Matter,''
  Phys.\ Rev.\ D {\bf 88}, 015004 (2013).
  %[arXiv:1303.7320 [hep-ph]].
  %%CITATION = ARXIV:1303.7320;%%

%\cite{Esmaili:2013gha}
\bibitem{Esmaili:2013gha} 
  A.~Esmaili and P.~D.~Serpico,
  %``Are IceCube neutrinos unveiling PeV-scale decaying dark matter?,''
  JCAP {\bf 1311}, 054 (2013).
  %[arXiv:1308.1105 [hep-ph]].
  %%CITATION = ARXIV:1308.1105;%%


%\cite{Assef:2010ew}
\bibitem{Assef:2010ew}
  R.~J.~Assef, C.~S.~Kochanek, M.~L.~N.~Ashby {\it et al.},
  %, C.~S.~Kochanek, M.~L.~N.~Ashby, M.~Brodwin, M.~J.~I.~Brown, R.~Cool, W.~Forman and A.~H.~Gonzalez {\it et al.},
  %``The Mid-IR and X-ray Selected QSO Luminosity Function,''
  Astrophys.\ J.\  {\bf 728}, 56 (2011).
  %[arXiv:1001.4529 [astro-ph.CO]].
  %%CITATION = ARXIV:1001.4529;%%

%\cite{Kistler:2009mv}
\bibitem{Kistler:2009mv} 
  M.~D.~Kistler, H.~Yuksel, J.~F.~Beacom, A.~M.~Hopkins and J.~S.~B.~Wyithe,
  %``The Star Formation Rate in the Reionization Era as Indicated by Gamma-ray Bursts,''
  Astrophys.\ J.\  {\bf 705}, L104 (2009).
  %[arXiv:0906.0590 [astro-ph.CO]].
  %%CITATION = ARXIV:0906.0590;%%

%\cite{Hopkins:2006bw}
\bibitem{Hopkins:2006bw} 
  A.~M.~Hopkins and J.~F.~Beacom,
  %``On the normalisation of the cosmic star formation history,''
  Astrophys.\ J.\  {\bf 651}, 142 (2006).
  %[astro-ph/0601463].
  %%CITATION = ASTRO-PH/0601463;%%

%\cite{Yuksel:2012zy}
\bibitem{Yuksel:2012zy} 
  H.~Yuksel and M.~D.~Kistler,
  %``The Cosmic MeV Neutrino Background as a Laboratory for Black Hole Formation,''
  arXiv:1212.4844.
  %%CITATION = ARXIV:1212.4844;%%

%\cite{Kistler:2013jza}
\bibitem{Kistler:2013jza} 
  M.~D.~Kistler, H.~Yuksel and A.~M.~Hopkins,
  %``The Cosmic Star Formation Rate from the Faintest Galaxies in the Unobservable Universe,''
  arXiv:1305.1630.
  %%CITATION = ARXIV:1305.1630;%%





\end{thebibliography}
\end{document}